\documentclass[10pt,twocolumn,letterpaper]{article}

\usepackage{iccv}
\usepackage{times}
\usepackage{epsfig}
\usepackage{graphicx}
\usepackage{amsmath}
\usepackage{amssymb}

\usepackage[nocompress]{cite}
\usepackage{url}
\usepackage{amsmath,amssymb,graphicx}

\usepackage{lipsum} 

\usepackage[switch]{lineno}
\usepackage{dcolumn}
\usepackage{bm}
\usepackage{color}
\usepackage{multirow}
\usepackage{algorithm,algorithmic,bm,color}

\usepackage[utf8]{inputenc}
\usepackage[T1]{fontenc}
\usepackage{mathptmx}
\usepackage{siunitx}
\usepackage{adjustbox}
\usepackage{diagbox}
\usepackage{diagbox}
\usepackage{xcolor,colortbl}
\definecolor{Gray}{gray}{0.85}
\usepackage{enumitem}
\usepackage{amsmath}
\usepackage[accsupp]{axessibility}

\newcolumntype{a}{>{\columncolor{Gray}}c}
\newcolumntype{b}{>{\columncolor{white}}c}

\newcommand{\argmin}{\arg\!\min}

\newcommand{\Amat}{{\boldsymbol A}}

\newcommand{\Dmat}{{\boldsymbol D}}

\newcommand{\Hmat}[0]{{{\boldsymbol H}}}
\newcommand{\Imat}{{\boldsymbol I}}

\newcommand{\Mmat}[0]{{{\boldsymbol M}}}

\newcommand{\Tmat}[0]{{{\boldsymbol T}}}

\newcommand{\Xmat}{{\boldsymbol X}}
\newcommand{\Ymat}[0]{{{\boldsymbol Y}}}
\newcommand{\Zmat}{{\boldsymbol Z}}

\newcommand{\bv}{\boldsymbol{b}}
\newcommand{\cv}{{\boldsymbol{c}}}

\newcommand{\ev}[0]{{\boldsymbol{e}}}

\newcommand{\uv}[0]{{\boldsymbol{u}}}
\newcommand{\vv}{\boldsymbol{v}}

\newcommand{\xv}{\boldsymbol{x}}
\newcommand{\yv}{\boldsymbol{y}}

\newcommand{\zv}{\boldsymbol{z}}

\newcommand{\Thetamat}{\boldsymbol{\Theta}}

\newcommand{\ts}{^{\top}}
\newcommand{\inv}{^{-1}}

\usepackage[pagebackref=true,breaklinks=true,letterpaper=true,colorlinks,bookmarks=false]{hyperref}

\iccvfinalcopy 

\def\iccvPaperID{8206} 

\ificcvfinal\fi

\begin{document}

\title{Self-supervised Neural Networks for Spectral Snapshot Compressive Imaging}

\author{Ziyi Meng\qquad Zhenming Yu\qquad Kun Xu\\
{Beijing University of Posts and Telecommunications}\\
{\tt\small \{mengziyi,yuzhenming,xukun\}@bupt.edu.cn}
\and
Xin Yuan$^*$\\
{ Westlake University}\\
{\tt\small xyuan@westlake.edu.cn}
}

\maketitle
\ificcvfinal\thispagestyle{empty}\fi

\begin{abstract}
We consider using {\bf\em untrained neural networks} to solve the reconstruction problem of snapshot compressive imaging (SCI), which uses a two-dimensional (2D) detector to capture a high-dimensional (usually 3D) data-cube in a compressed manner. 
Various SCI systems have been built in recent years to capture data such as high-speed videos, hyperspectral images, and the state-of-the-art reconstruction is obtained by the deep neural networks. 
However, most of these networks are trained in an end-to-end manner by a large amount of corpus with sometimes simulated ground truth, measurement pairs. In this paper, inspired by the untrained neural networks such as deep image priors (DIP) and deep decoders, we develop a framework by integrating DIP into the plug-and-play regime, leading to a self-supervised network for spectral SCI reconstruction.
Extensive synthetic and real data results show that the proposed algorithm without training is capable of achieving competitive results to the training based networks.  
Furthermore, by integrating the proposed method with a pre-trained deep denoising prior, we have achieved state-of-the-art results. {Our code is available at \url{https://github.com/mengziyi64/CASSI-Self-Supervised}.}
\end{abstract}

\let\thefootnote\relax\footnotetext{$^*$Corresponding author.} 
\section{Introduction}

\begin{figure}[htbp!]
\centering
{\includegraphics[width=.98\linewidth]{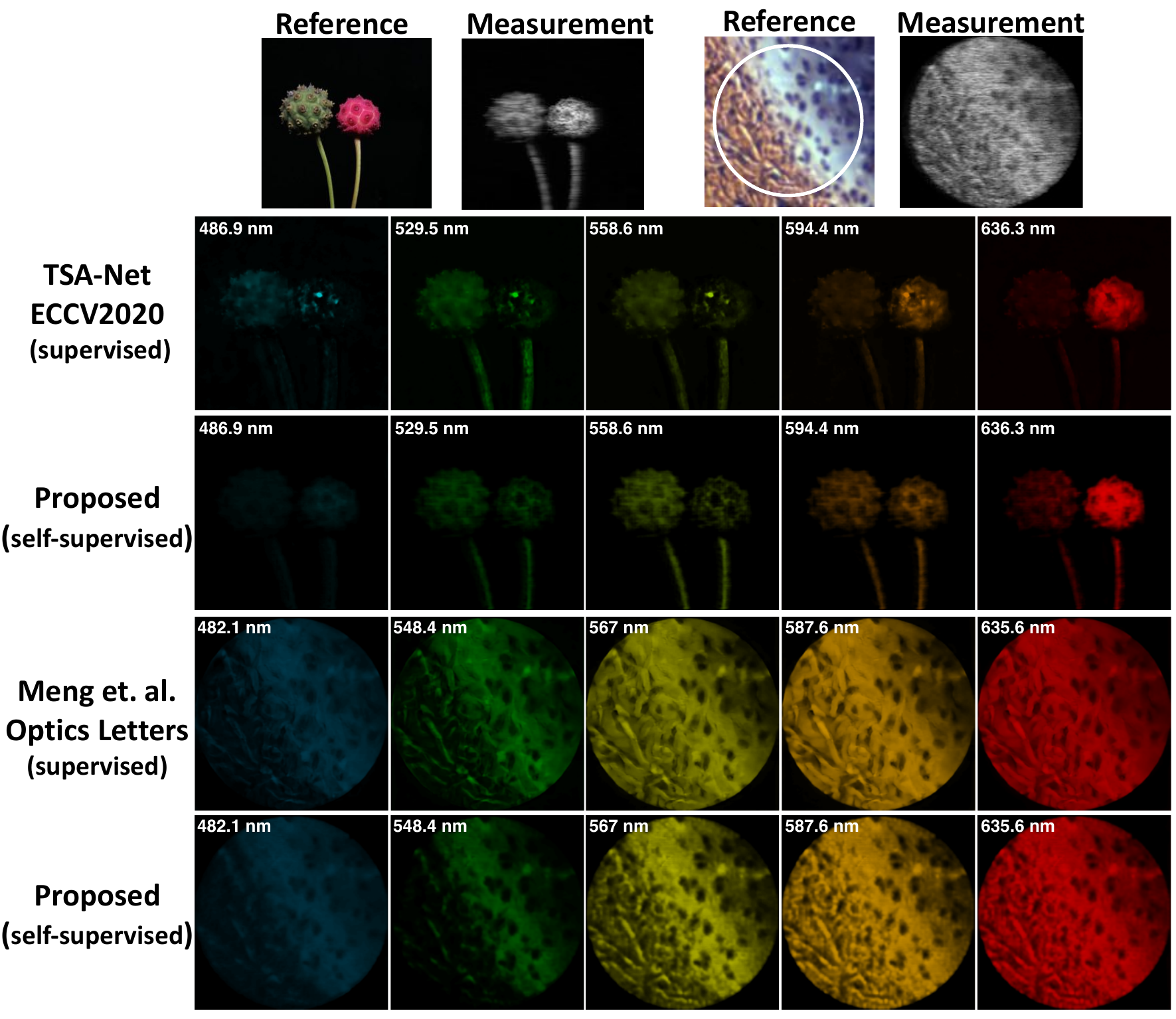}}
\vspace{-1mm}
\caption{Reconstructed {\bf real data} of {\em Plant} (upper) and {\em Dog olfactory membrane section} (lower), captured by the spectral SCI systems in~\cite{Meng20ECCV_TSAnet} and~\cite{meng2020snapshot}, respectively. We show 5 out of 28 (upper) and 5 out of 24 (lower) spectral channels  of the two scenes and compare our proposed self-supervised method (PnP-DIP that does not need training data) with two supervised algorithms (need training data), respectively.
}
\label{fig:real_1}
\vspace{-4mm}
\end{figure}

Recent advances in artificial intelligence and robotics have led to high demands to capture multi-dimensional high resolution data, such as high-speed videos, hyperspectral images, etc. 
This brings unprecedented challenges to existing imaging devices. On the other hand, compressive sensing (CS)~\cite{Candes06ITT,Donoho06ITT} has provided us an alternative way to devise imaging systems to capture these high-dimensional data.
As one representative technique based on CS, snapshot compressive imaging (SCI)~\cite{Liu18TPAMI,Yuan2021_SPM,Qiao2021_TCI_CSOCT} employs the multiplexing technique to impose the modulation in the optical path and captures the 3D spectral or temporal data-cube  using a 2D detector in a compressed way. 
An SCI system is thus composed of a hardware encoder to capture the compressed measurement and a software decoder to reconstruct the desired 3D data-cube.
In this paper, we consider the spectral SCI system, where the pioneering work is the coded aperture snapshot spectral imager (CASSI)~\cite{Gehm07,Wagadarikar08CASSI}, in which a physical mask and a prism are utilized to implement the multiplexing modulation. 
This work focuses on the algorithm design and thus the aforementioned software decoder in SCI. 
Specifically, we develop a reconstruction framework that integrates the {\em untrained neural networks} as priors~\cite{Ulyanov_2018_CVPR} into the plug-and-play (PnP) algorithms~\cite{Venkatakrishnan_13PnP,Sreehari16PnP}.

\subsection{Motivation}
It has been over a decade since the CASSI being built in the lab and one main bottleneck as in other CS systems was the reconstruction algorithm, which was usually based on iterative optimization.
These algorithms are either slow~\cite{Liu18TPAMI} or low quality~\cite{Bioucas-Dias2007TwIST}.
Thanks to deep learning, both the speed and quality have been improved significantly during the past few years as various deep networks~\cite{Wang19_CVPR_HSSP,Miao19ICCV,Wang_2020_CVPR,Meng20ECCV_TSAnet,Huang2021_CVPR_GSMSCI,Yuan18OE} have been built to implement the software decoder.
However, there is one challenge that needs to be addressed: {\em the training data}. 
It is well known that sufficient training data are very important to the performance of the deep networks. 
Unfortunately, for the hyperspectral imaging considered in this work, very limited data are available that can be used for training and most existing networks are still based on the CAVE~\cite{CAVE_spectraldata_07}, ICVL~\cite{arad_and_ben_shahar_2016_ECCV} and KAIST~\cite{Choi17TOG} data.  
Though good results have been obtained on synthetic and some real data, we have noticed that for some spectra that were not existed in the training data, these networks can not reconstruct the desired spectral cube well. {For example, the recovered results of $\lambda$-net~\cite{Miao19ICCV} in the {\em Bird} data~\cite{kittle2010multiframe} show big errors 
in the spectral accuracy due to the {\em mismatch} between the training data and the real data.}
The short of training data might have limited the generalizability of existing networks.

To address this challenge, one main breakthrough of using deep learning in inverse problem is the untrained neural networks such as the deep image prior (DIP)~\cite{Ulyanov_2018_CVPR} and deep decoder~\cite{heckel_deep_2018}, which utilize the neural networks to learn the priors from the {\rm raw measurements} directly and thus does not need any training data. 
Both DIP and deep decoder have shown promising results in some image restoration results such as image denoising, inpainting and super-resolution. 
{One straightforward way is to directly use these networks in the SCI reconstruction problem; however, after extensive experiments, we found it is difficult to obtain good results in this manner. Please refer to Table~\ref{Tab:SFvsDF} for a detailed comparison, where we call this direct usage as the ``sole DIP''.}
Since the goal of the untrained neural network is to learn a prior, in this work, we apply this prior (learning during reconstruction) into the recently advanced PnP framework~\cite{Venkatakrishnan_13PnP,Sreehari16PnP,yuan2020plug,PnP_SCI_arxiv2021}, with an optionally different kind of prior to solve the spectral SCI reconstruction. The network is optimized as the iteration in PnP going on during the reconstruction, and thus leading to a {\em self-supervised} deep learning framework. 

\subsection{Contributions}
The goal of this work is to develop a self-supervised neural network for the spectral SCI reconstruction, which enjoys the strong learning capability of deep networks but does not need any training data.
Specific contributions are:
\vspace{-2mm}
\begin{itemize}[leftmargin=*]
\setlength{\itemsep}{0pt}
\setlength{\parsep}{0pt}
\setlength{\parskip}{0pt}
\item A self-supervised framework is proposed for spectral SCI reconstruction.
\item An alternating optimization algorithm is developed to solve the joint network learning and reconstruction. 
\item Extensive results on both synthetic and real datasets verify the superiority of the proposed approach. 
\item Our proposed algorithm is robust to the {\em Poisson} noise, which happens in real measurements.
\end{itemize}
\vspace{-2mm}
Importantly, our model does not need any training data, but with fine tuning on parameters for each dataset, competitive results are obtained with similar quality to the recently proposed supervised deep learning algorithms. 
Please refer to Fig.~\ref{fig:real_1} for two {\bf real data} results captured by two different spectral SCI systems. 
Furthermore, by integrating our proposed approach with the pre-trained HSI deep denoising prior~\cite{Zheng20_PRJ_PnP-CASSI}, we have achieved state-of-the-art results. 

\subsection{Related Work}
In the past decade, spectral SCI systems have been developed by various hardware designs~\cite{Wagadarikar08CASSI,Lin14ToG,Yuan15JSTSP,wu2011development,Ma2021_LeSTI_OE}. 
For the reconstruction, since the inverse problem is ill-posed, regularizers or priors are widely used, such as the sparsity~\cite{Figueiredo07_GPSR} and total variation (TV)~\cite{Bioucas-Dias2007TwIST}. 
Later, the patch-based methods such as dictionary learning~\cite{Aharon06TSP,Yuan15JSTSP} and Gaussian mixture models~\cite{Yang14GMMonline}, and group sparsity~\cite{Wang17_PAMI_GSC} and low-rank models~\cite{Liu18TPAMI,He2021_TIP} have been developed. 
The main bottleneck of these iterative optimization-based algorithms is the low reconstruction speed, especially for the large-scale dataset. 
Another limitation is that these handcrafted priors may not fit every data. 
 %

Inspired by the high performance of deep learning for other inverse problems~\cite{Barbastathis:19,Yuan18OE}, convolutional neural networks (CNN) have been used to solve the inverse problem of spectral SCI for the sake of high speed~\cite{meng2020snapshot,Miao19ICCV,Wang19_CVPR_HSSP,Meng20ECCV_TSAnet,Meng2021_ICIP_Ploss}.
These networks (trained in a supervised manner) have led to better results than the optimization counterparts, given sufficient training data and time, which usually take days or weeks.  
After training, the network can output the reconstruction instantaneously and thus lead to an end-to-end spectral SCI sampling and reconstruction system~\cite{Meng20ECCV_TSAnet}.
However, these networks are usually system specific. For example, different numbers of spectral channels exist in different systems. 
Further, due to the different designs of the mask (modulation patterns), the trained CNNs cannot be used in other systems, while re-training a new network from scratch would take a long time.

Therefore, {\em efficient and effective unsupervised} algorithms are still highly desired as researchers are eager to verify the system when a new hardware is built. Unfortunately, the two classes of algorithms cannot fulfill this basic requirement.
Thanks to the untrained neural networks being proposed for inverse problems~\cite{Ulyanov_2018_CVPR,Qiao2021_MicroCACTI}, we now can develop {\em a new class of algorithms that enjoys the power of deep neural networks but does not require any training data}.  

In~\cite{zhang2019hyperspectral}, DIP is employed as a refinement process of the trained network for the reconstruction of a single image. This is very different from our proposed self-supervised method. 
The other related work is DeepRED~\cite{Mataev_2019_ICCVW_DeepRed}, where DIP is combined with Regularization by Denoising (RED)~\cite{Romano17RED_SIAM} and achieved better results than DIP itself.
Our work differentiates from DeepRED and the follow up work regularization
by artifact-removal (RARE)~\cite{Liu20_JSTSP_RARE} in the following perspectives: $i$) instead of using the RED prior, we combine DIP with implicit conventional priors, where any existing denoiser can be used; $ii$) {during deriving the solution of our proposed model, we utilized two fidelity terms ($\|\yv-\Hmat\xv\|_2^2$ and $\|\yv-\Hmat\Tmat_{\Thetamat}(\ev)\|_2^2$ in Eq.~\eqref{eq:xTheta_joint3}) to ensure both priors leading to the same result, while only one fidelity term was used in DeepRED (only $\|\yv-\Hmat\Tmat_{\Thetamat}(\ev)\|_2^2$ in Eq.~\eqref{eq:xTheta_joint1}) and we have experimentally shown that our proposed method leads to better results}; $iii$) we apply the proposed method to the spectral SCI reconstruction of  both synthetic and real data, which is different from the tasks considered in DeepRED.
Most recently, a pre-trained hyperspectral images (HSI) deep denoising prior~\cite{Zheng20_PRJ_PnP-CASSI} has been used in PnP for spectral SCI reconstruction. This is another way of using neural networks.


\section{Spectral SCI System \label{Sec:SCI_model}}
\begin{figure}[htbp!]
\vspace{-6mm}
\centering
{\includegraphics[width=.95\linewidth]{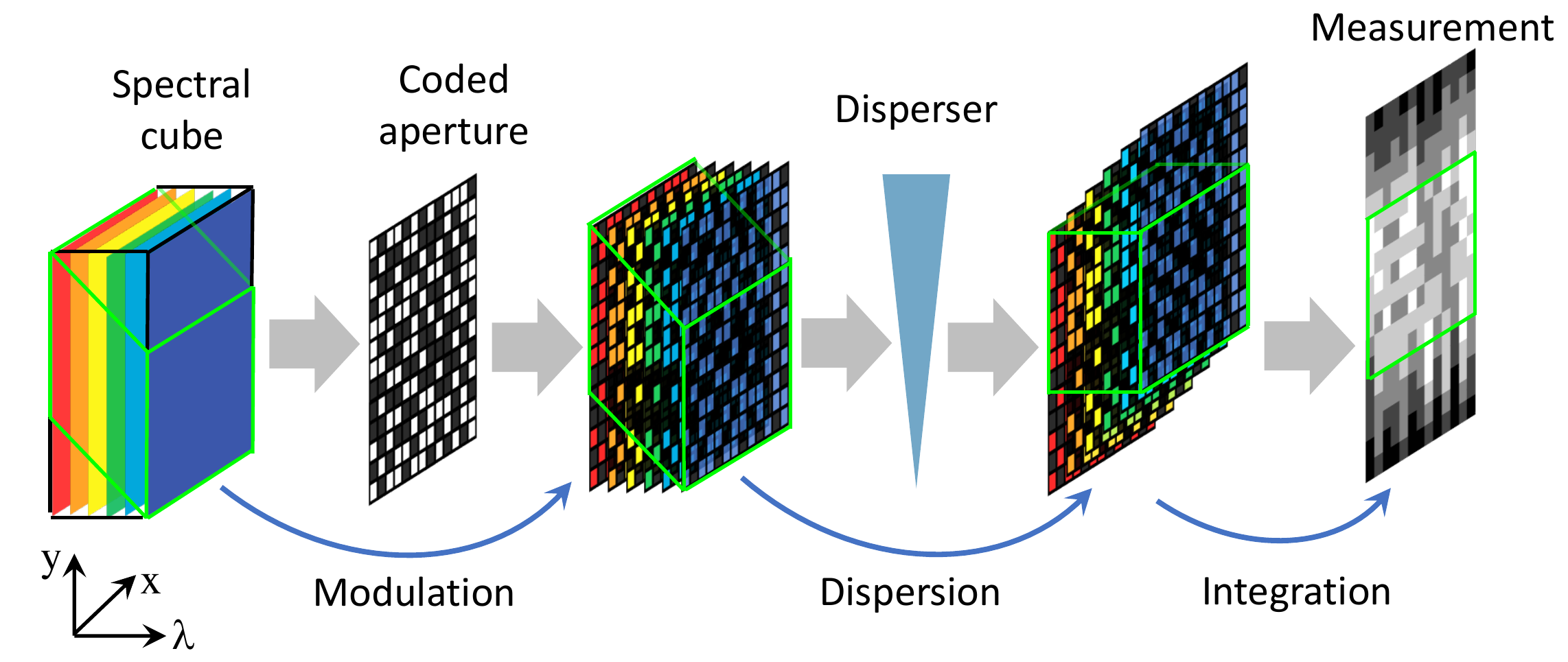}}
\vspace{-1mm}
\caption{Schematic diagrams of spectral SCI, a.k.a., coded aperture snapshot spectral imaging (CASSI) system.}
\vspace{-2mm}
\label{fig:CASSI}
\end{figure}
The underlying principle of SCI system is to encode the high-dimensional data  onto a 2D measurement. As one of the earliest proposed SCI systems, CASSI system~\cite{Wagadarikar08CASSI} captures the spectral image cube in a snapshot using simple and low cost hardware. Fig. \ref{fig:CASSI} shows a schematic diagram of CASSI. The spectral image data-cube is first modulated by a coded aperture (\ie, a fixed mask), and then the coded data-cube is spectrally dispersed by the dispersing element, and finally integrated across the spectral dimension to a 2D measurement captured by the camera sensor.

Recalling Fig.~\ref{fig:CASSI}, let $\Xmat^0 \in {\mathbb R}^{n_x \times n_y \times n_\lambda}$ denote the spatio-spectral data-cube to be captured, which is first modulated by the mask $\Mmat\in {\mathbb R}^{n_x \times n_y}$, \ie, for $m = 1,\dots, n_{\lambda}$, we have
\begin{equation}
{\Xmat}' (:,:,m) = {\Xmat} (:,:,m) \odot \Mmat,
\end{equation} 
where ${\Xmat}' \in {\mathbb R}^{n_x \times n_y \times n_\lambda}$  is the modulated cube, $\odot$ represents the element-wise multiplication and ${\Xmat} (:,:,m)$ denotes the $m$-th channel in the spectral cube of $\Xmat$.
After passing the disperser, the modulated cube $\Xmat'$ is tilted, \ie, each spectral channel is shifted spatially on the dispersion direction ($y$-axis in Fig.~\ref{fig:CASSI}).
Let $\Xmat''\in {\mathbb R}^{n_x \times (n_y + n_{\lambda}-1) \times n_{\lambda}}$ denote the tilted cube, and we have 
$\Xmat'' (u,v, m) = \Xmat'(x, y + d(\lambda_m - \lambda_c), m)$,
where $(u,v)$ indicates the coordinate system on the detector plane, $\lambda_m$ is the wavelength at $m$-th channel and $\lambda_c$ denotes the center-wavelength that does not shift direction after disperser. Then, $d(\lambda_m -\lambda_c)$ signifies the spatial shifting  for the $m$-th spectral channel.

Finally, the 2D compressed measurement on the sensor plane $y(u,v)$ is acquired by the integration on the designed wavelength range $[\lambda_{\rm min}, \lambda_{\rm max}]$, and thus can be expressed by
\begin{equation}
\textstyle{ \Ymat = \sum_{m=1}^{n_{\lambda}}  \Xmat'' (:,:, m) + \Zmat}.
\label{Eq:Sensing}
\end{equation}
In other words, $\Ymat \in {\mathbb R}^{n_x \times (n_y + n_{\lambda}-1)}$ is a {\em compressed} image which is formed by a function of the desired information corrupted by the measurement noise $\Zmat\in {\mathbb R}^{n_x \times (n_y + n_{\lambda}-1)}$. 

We further give the vectorized formulation of this process. 
Let ${\rm vec}(\cdot)$  denote the matrix vectorization operation, \ie, concatenating columns into one vector. 
Then, we define
$\yv = {\rm vec}(\Ymat)$, 
$\zv = {\rm vec}(\Zmat) \in {{\mathbb R}^{n_x(n_y+n_\lambda-1)}}$ and
$\xv = [\xv_1\ts, \dots,\xv_{n_\lambda}\ts]\ts$,
where, for 
$m=1,\dots, n_\lambda$, $\xv_m = {\rm vec}( \Xmat (:,:, m))$.
In addition, we define the sensing matrix as
\begin{equation}
\Hmat = \left[\Dmat_1, \dots, \Dmat_{n_{\lambda}}\right] \in   {{\mathbb R}^{n_x(n_y+n_{\lambda}-1) \times  n_x n_y n_{\lambda}}}, \label{Eq:Phimat}
\end{equation}
where $\Dmat_{m} = \left[\begin{array}{c}
{\bf 0}^{(1)}\\ 
\Amat_{m}\\ 
{\bf 0}^{(2)}
\end{array}\right]\in {\mathbb R}^{n_x(n_y+n_{\lambda}-1) \times  n_x n_y}$
with
$\Amat_{m} = {\rm Diag} ({\rm vec}(\Mmat))\in {\mathbb R}^{n_x n_y \times  n_x n_y}$ being a diagonal matrix with ${\rm vec}(\Mmat)$ as its diagonal elements, and ${\bf 0}^{(1)}\in {\mathbb R}^{(m-1) \times  n_x n_y}$, ${\bf 0}^{(2)}\in {\mathbb R}^{(n_{\lambda}-m) \times  n_x n_y}$ are zero matrices.
As such, we then can rewrite the matrix formulation of \eqref{Eq:Sensing} as
\begin{equation}
\yv = \Hmat {\xv} + \zv. \label{Eq:CS_forwared}
\end{equation}
This model is similar to CS. However, due to the special structure of the sensing matrix $\Hmat$, most theories developed for CS cannot fit in this application. 
It has been proven that the signal can still be recovered even when $n_\lambda > 1$~\cite{Jalali19TIT_SCI}.

After capturing the measurement, the following task is given $\yv$ (captured by the camera) and $\Hmat$ (calibrated based on pre-designed hardware), solving $\xv$.

\section{Methods \label{Sec:methods}}


To recover the images from~\eqref{Eq:CS_forwared}, optimization algorithms usually employ a regularization term, or a prior, to confine the solution to the desired signal space. Let $R(\xv)$ denote the regularization term (prior) to be used, the reconstruction target is formulated as
\begin{equation}
	\textstyle{\hat{\xv} = \arg\min_{\xv} \frac{1}{2} \|\yv - \Hmat\xv\|_2^2 + \lambda  R(\xv)\,.}
\label{eq:map_reg}
\end{equation}
In the literature, different priors have been used for spectral SCI including TV~\cite{Bioucas-Dias2007TwIST}, sparsity~\cite{Figueiredo07_GPSR} and low-rank models~\cite{Liu18TPAMI}. 
However, these are all hand-crafted priors which might not fit all the experimental data.
On the other hand, recent researches have shown that deep neural networks are capable to learn complicated structures in the data, and more specifically, from the {\em corrupted measurement itself}. 

\subsection{Deep Image Prior (DIP) \label{Sec:sub_DIP}}
Starting from \eqref{eq:map_reg}, but removing the regularization term $R(\xv)$, deep image prior~\cite{Ulyanov_2018_CVPR} assumes that the desired signal $\xv$ is the output of a neural network, $\Tmat_{\Thetamat}(\ev)$, where $\ev$ is a random vector and $\Thetamat$ is the network's parameters to be learned.
Thereby, DIP suggests to solve
\begin{equation}
    \textstyle{\min_{\Thetamat} \|\yv - \Hmat \Tmat_{\Thetamat}(\ev)\|_2^2,}
\end{equation}
and the desired reconstruction will be $\hat{\xv} = \Tmat_{\Thetamat}(\ev)$. 
Note the key difference between DIP and other existing deep neural networks is that here $\Thetamat$ is specific for each measurement $\yv$. In fact,  $\Thetamat$ is learned from $\yv$.
By contrast, existing {supervised} networks learned the network parameters from the training data and are then fixed during testing (inference). 
In DIP, the training of $\Thetamat$ itself is the reconstruction process of $\xv$. 
This procedure is  thus {\em unsupervised} in the sense that no ground truth is used during 
the learning (meanwhile reconstruction). 
Over-fitting is avoided due to the implicit regularization imposed by the network and early stopping.

DIP indeed achieved good recovery results in some image restoration tasks such as image denoising, inpainting and super-resolution. 
However, for the challenging case of SCI considered here, directly applying DIP usually cannot give us good results since it is too ill-posed. For example, since the third (spectral) dimension of the data-cube is smashed into the single 2D measurement, using the DIP itself cannot recover the spectral information of the data-cube though sometimes, it can provide good spatially visual images.
By contrast, traditional priors such as TV can usually lead to a good spectral recovery but losing some spatial details. 
Therefore, in this work, we propose to use DIP as a ``prior'' and by incorporating it with other traditional priors, we arrive at a ``self-supervised''
framework for SCI reconstruction. 
On one hand, these two priors will compete with each other during reconstruction; on the other hand, they are also complementary with each other.
In other words, they will drag each other to avoid the other one sticking to a local minimum.

\subsection{Proposed Joint Framework \label{Sec:sub_pro}}
As mentioned above, we impose two priors, DIP + $R(\xv)$, on the spectral data-cube to be reconstructed.
This leads to the following formulation.
\begin{equation}
    \textstyle{(\hat\xv, \hat\Thetamat) = \argmin_{\xv, \Thetamat} \frac{1}{2}\|\yv - \Hmat\xv\|_2^2 + \lambda R(\xv), \text{~s.t.~} \xv = \Tmat_{\Thetamat}(\ev)}. \label{eq:xTheta_joint}
\end{equation}
As mentioned in~\cite{Mataev_2019_ICCVW_DeepRed}, though it looks simpler to only optimize $\Thetamat$ in \eqref{eq:xTheta_joint}, it is almost impossible to solve it directly. 
By introducing an auxiliary variable $\bv\in {\mathbb R}^{n_x n_y n_\lambda}$ and a balance parameter $\mu$, we aim to minimize
\begin{equation}
        \textstyle{\min_{\xv, \Thetamat}\frac{1}{2}\|\yv - \Hmat\Tmat_{\Thetamat}(\ev)\|_2^2 + \lambda R(\xv) + \mu \|\xv - \Tmat_{\Thetamat}(\ev) -\bv\|_2^2}. \label{eq:xTheta_joint1}
\end{equation}
Using the alternating direction method of multipliers (ADMM)~\cite{Boyd11ADMM}, the solution can be derived by splitting it into three subproblems. Similar derivations can be found in~\cite{Mataev_2019_ICCVW_DeepRed}. 
We show the derivation details in the supplementary materials (SM) and compare this with our proposed approach derived as follows in Table~\ref{Tab:SFvsDF}.

Due to the two priors being used in Eq.~\eqref{eq:xTheta_joint}, we find in the experiments that since we only enforce the results of DIP, thus $\Tmat_{\Thetamat}(\ev)$ close to the measurement $\yv$, which is the only available input to the algorithm, is not capable of merging the wellness of both priors. Therefore, in the following, we propose to minimize 
\begin{equation}
\begin{aligned}
     {\min_{\xv, \Thetamat} \frac{1}{2}\|\yv - \Hmat\xv\|_2^2 + \lambda R(\xv) + \frac{\rho}{2}\|\yv - \Hmat\Tmat_{\Thetamat}(\ev)\|_2^2},\\ \text{~ s.t.~} \xv = \Tmat_{\Thetamat}(\ev). \label{eq:xTheta_joint2}
    \end{aligned}
\end{equation}
In order to solve \eqref{eq:xTheta_joint2}, similarly, we introduce an auxiliary variable $\bv\in {\mathbb R}^{n n_\lambda}$ and the balance parameter $\mu$ and now aim to minimize
\begin{equation}
\begin{aligned}
  &(\hat\xv, \hat\Thetamat,\hat \bv) = \argmin_{\xv, \Thetamat,\bv} \textstyle{\frac{1}{2}\|\yv - \Hmat\xv\|_2^2 + \lambda R(\xv)} -  \frac{\mu}{2}\|\bv\|_2^2\\
	&+\textstyle{  \frac{\rho}{2}\|\yv - \Hmat \Tmat_{\Thetamat}(\ev)\|_2^2+  \frac{\mu}{2} \|\xv - \Tmat_{\Thetamat}(\ev) - \bv\|_2^2. }\label{eq:xTheta_joint3}   
\end{aligned}
\end{equation}
We  solve \eqref{eq:xTheta_joint3} iteratively composed of the following subproblems. In the derivation below, we use the superscript $k$ to denote the iteration number and for simplicity, we discard this index in some subproblems such as $\Thetamat$ and $\xv$. 
\begin{itemize}
	\item[1)] $\Thetamat$-subproblem:
	Given $\xv$ and $\bv$, we aim to solve ${\Thetamat}$ by
	\begin{equation}
		\textstyle{\hat\Thetamat = \argmin_{\Thetamat} \frac{\rho}{2}\|\yv-\Hmat \Tmat_{\Thetamat}(\ev)\|_2^2 +  \frac{\mu}{2} \|\xv - \Tmat_{\Thetamat}(\ev) -\bv\|_2^2}, \label{eq:thetamat_2term}
	\end{equation}
	which shares the similar spirit to the optimization done in DIP using back-propagation, modified by a proximity regularization that forces $\Tmat_{\Thetamat}(\ev)$ to be close to $\xv-\bv$. This proximity
term provides an additional stabilizing effect to the DIP minimization. 
For instance, in the U-net being used in our implementation, in the loss function, instead of only minimizing the first term in \eqref{eq:thetamat_2term} as in the DIP, we hereby used both terms as the loss function.  
This learned $\Tmat_{\Thetamat}(\ev)$ is thus playing the role of two-fold: i) denoising $\xv-\bv$, and ii) minimizing the measurement loss $\yv - \Hmat \Tmat_{\Thetamat}(\ev)$. $\mu$ and $\rho$ in \eqref{eq:thetamat_2term} are parameters to balance these two terms.

	\item[2)] $\xv$-subproblem: we aim to solve
	\begin{equation}
		\textstyle{\hat\xv = \argmin_{\xv} \frac{1}{2}\|\yv - \Hmat\xv\|_2^2 + \lambda R(\xv) + \frac{\mu}{2} \|\xv - \Tmat_{\Thetamat}(\ev) - \bv\|_2^2}. \nonumber
	\end{equation}
	Due to the three coupled terms and the implicit formulation of $R(\xv)$, we apply ADMM again here by introducing $\uv, \vv$.
	This leads to minimize
	\begin{align}
		\textstyle{\min_{\xv,\uv}} &	\textstyle{\frac{1}{2}\|\yv - \Hmat\xv\|_2^2 + \lambda R(\uv) 
		+ \frac{\mu}{2} \|\xv - \Tmat_{\Thetamat}(\ev) - \bv\|_2^2} \nonumber\\
		&\text{~s.~t.~ } \uv = \xv. \label{eq:xR_joint}
	\end{align}
     Eq.~\eqref{eq:xR_joint} is re-formulated as
     \begin{align}
     \textstyle{\min_{\xv,\uv} \frac{1}{2}\|\yv - \Hmat\xv\|_2^2} &+ \lambda R(\uv) +  \textstyle{\frac{\mu}{2} \|\xv - \Tmat_{\Thetamat}(\ev) - \bv\|_2^2 } \nonumber \\
     	&+  \textstyle{\frac{\eta}{2} \|\xv - \uv - \vv\|_2^2 -\frac{\eta}{2} \|\vv\|_2^2}. \label{eq:xR_joint2}
     \end{align}
   This is solved by the following sub-problems:
   \begin{list}{\labelitemi}{\leftmargin=10pt \topsep=0pt \parsep=0pt}
   	\item[2.1)] $\xv$-subproblem:
   	\begin{equation}
   	\begin{aligned}
   		\textstyle{\hat\xv = \argmin_{\xv} \frac{1}{2}\|\yv - \Hmat\xv\|_2^2} &+ \textstyle{\frac{\mu}{2}\|\xv - \Tmat_{\Thetamat}(\ev) - \bv\|_2^2}  \\
   		&+  \textstyle{\frac{\eta}{2} \|\xv - \uv - \vv\|_2^2}. \label{eq:xv_3term}
   	\end{aligned}
   	\end{equation}
   This is a quadratic form and due to the special structure of $\Hmat$, it has a closed-form solution 
       $\hat\xv = (\Hmat\ts\Hmat + \mu \Imat + \eta\Imat)\inv  [\Hmat\ts\yv + \mu(\Tmat_{\Thetamat}(\ev) + \bv) + \eta (\uv+\vv)]$. 
   Recalling $\Hmat$ in Eq.~\eqref{Eq:Phimat}, we can observe that $\Hmat\Hmat\ts$ is a diagonal matrix. 
   Using the the matrix inversion lemma (Woodbury matrix identity)~\cite{hager1989updating}:
   $(\Hmat\ts\Hmat + \mu \Imat + \eta\Imat)\inv = (\mu + \eta)\inv - (\mu + \eta)\inv \Hmat\ts (\Imat + (\mu+\eta)\Hmat\Hmat\ts)\inv \Hmat (\mu + \eta)\inv$,
   the solution of $\hat\xv$ can be obtained efficiently by
%
\begin{align}
\cv &\stackrel{\rm def}{=} (\mu(\Tmat_{\Thetamat}(\ev) + \bv)  + \eta (\uv + \vv))/({\mu + \eta}),\nonumber\\
 \hat\xv& =  \textstyle{ {\cv}  + \Hmat\ts (\yv - \Hmat\cv ) \oslash ({\rm Diag}(\Hmat\Hmat\ts) + \mu + \eta)}, 
    \label{Eq:x_final}
\end{align}

where ${\rm Diag}(~)$ extracts the diagonal elements of the ensued matrix and $\oslash$ denotes the element-wise division.
   \item[2.2)] $\uv$-subproblem:
   	$\hat\uv = \textstyle{\argmin_{\uv}  \eta \|\xv - \uv - \vv\|_2^2 + \lambda R(\uv)}$. 
   This is a denoising problem and depending on the selection of $R$, we have
   \begin{equation}
   	\quad \hat\uv = {\cal D}_\sigma (\xv - \vv), \label{eq:uv_update} 
   	\end{equation}
    where $\sigma$ is the estimated noise level depending on $\lambda/\eta$.
    \item[2.3)] $\vv$ is updated by 
	\begin{equation}
		\quad \vv^{k+1} = \vv^{k+1} - (\xv^k - \uv^k),\label{eq:vv_update}
	\end{equation}
	where $k$ denotes the iteration number.
   \end{list}
   \item[3)] $\bv$ is updated  by
	\begin{equation}
		\qquad 		\bv^{k+1} =  \bv^k -(\xv^k - \Tmat_{\Thetamat^k}(\ev)). \label{eq:bv_update}
	\end{equation}
\end{itemize}

We change the orders of updating parameters from $\xv$, $\uv$, $\vv$, to $\Thetamat$ and lastly $\bv$ in our experiments and the entire algorithm is exhibited in Algorithm \ref{algo:PnP_DIP}.

\begin{table*}
\caption{PSNR in dB and SSIM reconstructed by different algorithms on 10 synthetic data.}
\begin{center}
\resizebox{.95\textwidth}{!}{
\begin{tabular}{b|b|b|b|b|b|b|a|a}
\hline
Algorithms & TwIST~\cite{Bioucas-Dias2007TwIST} & ADMM-TV~\cite{Yuan16ICIP_GAP} & DeSCI~\cite{Liu18TPAMI} & PnP-HSI~\cite{Zheng20_PRJ_PnP-CASSI} & DeepRED~\cite{Mataev_2019_ICCVW_DeepRed} & TSA-Net~\cite{Meng20ECCV_TSAnet} & \shortstack{PnP-DIP\\ (Proposed)} & \shortstack{PnP-DIP-HSI\\ (Proposed)}\\
\hline\hline
Scene 1 & 24.62, 0.714 & 25.77, 0.729 & 27.15, 0.794 & 26.35, 0.712 & 28.27, 0.769 & 31.26, 0.887 & 31.98, 0.862 & {\bf 32.70}, {\bf 0.898} \\
\hline
Scene 2 & 20.47, 0.578 & 21.39, 0.589 & 22.26, 0.694 & 22.60, 0.613 & 21.64, 0.602 & 26.88, {\bf 0.855} & 26.57, 0.767 & {\bf 27.27}, 0.832 \\
\hline
Scene 3 & 21.12, 0.746 & 23.14, 0.737 & 26.56, 0.877 & 26.78, 0.786 & 24.42, 0.769 & 30.03, {\bf 0.921} & 30.37, 0.862 & {\bf 31.32}, 0.920  \\
\hline
Scene 4 & 34.20, 0.907 & 33.70, 0.834 & 39.00, 0.965 & 37.61, 0.877 & 37.93, 0.927 & 39.90, 0.964 & 38.71, 0.930 & {\bf 40.79}, {\bf 0.970} \\
\hline
Scene 5 & 22.13, 0.688 & 23.43, 0.699 & 24.80, 0.778 & 24.88, 0.721 & 25.04, 0.757 & 28.89, 0.878 & 29.09, 0.849 & {\bf 29.81}, {\bf 0.903} \\
\hline
Scene 6 & 22.67, 0.696 & 23.68, 0.648 & 23.55, 0.753 & 24.85, 0.685 & 26.14, 0.743 & {\bf 31.30}, {\bf 0.895} & 29.85, 0.848 & 30.41, 0.890 \\
\hline
Scene 7 & 17.57, 0.603 & 18.62, 0.603 & 20.03, 0.772 & 20.12, 0.648 & 22.62, 0.777 & 25.16, 0.887 & 27.69, 0.864 & {\bf 28.18}, {\bf 0.913} \\
\hline
Scene 8 & 22.73, 0.702 & 23.39, 0.631 & 20.29, 0.740 & 23.80, 0.691 & 23.42, 0.674 & {\bf 29.69}, {\bf 0.887} & 28.96, 0.843 & 29.45, 0.885 \\
\hline
Scene 9 & 22.60, 0.733 & 23.25, 0.682 & 23.98, 0.818 & 25.11, 0.687 & 28.35, 0.840 & 30.03, 0.903 & 33.55, 0.881 & {\bf 34.55}, {\bf 0.932} \\
\hline
Scene 10 & 23.52, 0.610 & 23.86, 0.559 & 25.94, 0.666 & 24.57, 0.611 & 25.62, 0.723 & 28.32, 0.848 & 28.05, 0.833 & {\bf 28.52}, {\bf 0.863} \\
\hline
Average & 23.16, 0.697 & 24.02, 0.671 & 25.86, 0.785 & 25.67, 0.703 & 26.35, 0.758 & 30.15, 0.893 & 30.48, 0.854 & {\bf 31.30}, {\bf 0.901} \\
\hline
\end{tabular}
}
\end{center}
\label{Tab:simu}
\vspace{-7mm}
\end{table*}

Note that if we set $\lambda=0$, this means only the DIP is used in our proposed algorithm (this also leads to $\eta=0$). However, this is different from directly using DIP in the SCI problem due to the two $\ell_2$ fidelity terms in Eq.~\eqref{eq:xTheta_joint2}, which connects the closed-form projection of $\xv$ and DIP and then initiates the iterations.
This will avoid the local minimum that DIP usually sticks in. 
In our derivation, we did not impose the explicit priors of $R(\xv)$ such as the TV or sparsity with the following considerations:
\vspace{-2mm}
\begin{itemize}[leftmargin=*]
\setlength{\itemsep}{0pt}
\setlength{\parsep}{0pt}
\setlength{\parskip}{0pt}
    \item In {\bf real data} captured by the CASSI systems, usually a TV based algorithm can lead to a good initialization, which can be used a {\em warm} starting point of CNN based algorithms and also our proposed algorithm. This has also been used in previous algorithms~\cite{Liu18TPAMI,Zheng20_PRJ_PnP-CASSI}.
    \item When a deep hyperspectral images denoiser is trained~\cite{Zheng20_PRJ_PnP-CASSI}, we can use it jointly with our proposed approach. This will boost up the results and in fact, by integrating our proposed approach with the  HSI deep denoising prior trained in~\cite{Zheng20_PRJ_PnP-CASSI}, we can obtain the state-of-the-art result, which is of higher quality than either one alone (DIP or PnP-HSI).
\end{itemize}

\vspace{-5mm}
{\setlength{\textfloatsep}{0pt}
\begin{algorithm}[!htbp]
	\caption{Self-supervised algorithm for spectral SCI}
	\begin{algorithmic}[1]
		\REQUIRE$\Hmat$, $\yv$.
		\STATE Initial $\Thetamat,\bv,\uv,\vv $ and $\mu, \eta$.
		\WHILE{Not Converge}
		\STATE Update $\xv$ by Eq.~\eqref{Eq:x_final}. 
		\STATE Update $\uv$ by denoiser by Eq.~\eqref{eq:uv_update}.
		\STATE Update $\vv$ by Eq.~\eqref{eq:vv_update}.
		\STATE Update $\Thetamat$ by DIP with two loss terms in Eq.~\eqref{eq:thetamat_2term}.
		\STATE Update $\bv$ by Eq.~\eqref{eq:bv_update}.
		\ENDWHILE
	\end{algorithmic}
	\label{algo:PnP_DIP}
\end{algorithm}}
\vspace{-2mm}

\section{Results}
In this section, we validate the proposed self-supervised algorithm PnP-DIP and the boosted version of PnP-DIP-HSI on synthetic datasets and real data, and compare them with other iterative algorithms and supervised deep learning methods for spectral SCI reconstruction.
\subsection{Results on Synthetic Data}
\noindent \textbf{Implementation Details.}
For the implementation of DIP, we use U-net~\cite{Unet_RFB15a} as the self-supervised neural network. We discarded the skip connections in the U-net as suggested in the DIP paper~\cite{Ulyanov_2018_CVPR}. The network input $\ev$ is a random vector with the same size as the signal $\xv$ to be recovered, and we keep the vector fixed in each ADMM iteration for a fixed task (corresponding to one compressed measurement). Early stopping is used to avoid the over-fitting. Specifically, we use early stopping earlier in the first few ADMM iterations, and then increase the DIP iterations gradually in the later ADMM iterations. This is reasonable due to the improvement of the image quality as the increase of the ADMM iterations (outer-loop in Algorithm~\ref{algo:PnP_DIP}). Note that the network parameter $\Thetamat$ is set to zero after the DIP process in each ADMM iteration; this means $\Thetamat$ is re-trained from beginning in each iteration.
This avoids the local minimum that DIP stuck in the last iteration. 
For the loss function of DIP \eqref{eq:thetamat_2term}, the balance parameter $\rho$ and $\mu$ is set to $\rho/\mu = 0.1$, \ie, we use a smaller weight for the measurement loss $\yv - \Hmat \Tmat_{\Thetamat}(\ev)$. This ensures the stabilization of the DIP minimization. We use Adam~\cite{kingma2014adam} as the optimizer and the learning rate is set to be 0.001.

\begin{figure}[htbp!]
\vspace{-1mm}
\centering
{\includegraphics[width=.98\linewidth]{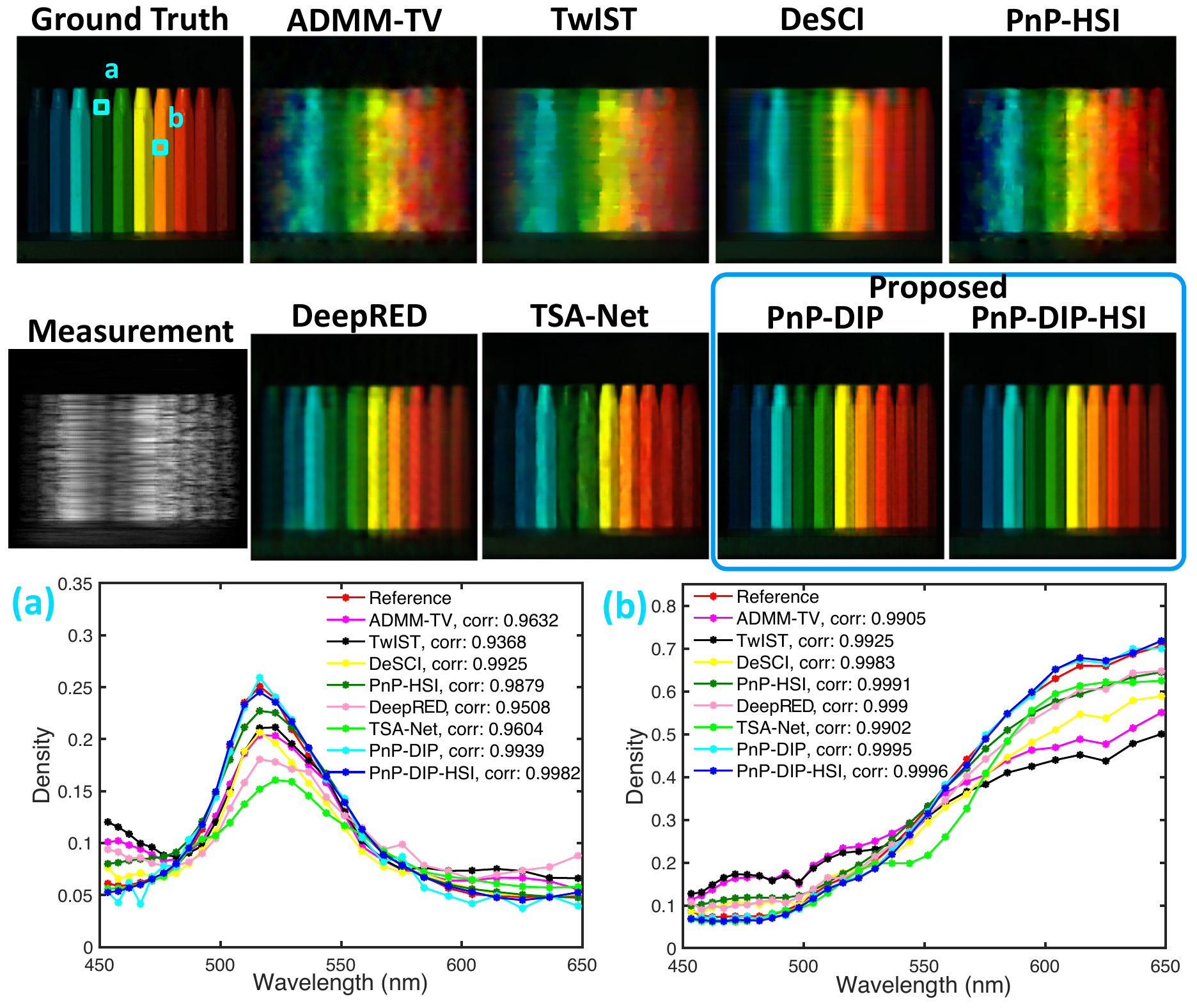}}
\vspace{-1mm}
\caption{Reconstructed {\bf synthetic data} (sRGB) of {\em Scene 9} by 8 algorithms. We show the reconstructed spectral curves on selected regions to compare the spectral accuracy of different algorithms.
}
\label{fig:simu_1}
\vspace{-5mm}
\end{figure}

For the CASSI reconstruction, following Algorithm \ref{algo:PnP_DIP}, we propose two methods, PnP-DIP and PnP-DIP-HSI. 
As mentioned before, PnP-DIP only employs DIP as the prior, \ie, we set $\lambda, \eta = 0$. Other parameters are initialized by \{$\Tmat_{\Thetamat}(\ev)=\Hmat\ts\yv$, $\bv=0$\}, and $\mu=0.01$, In  PnP-DIP-HSI, a trained HSI denoiser~\cite{Zheng20_PRJ_PnP-CASSI} is combined with the DIP in the last few ADMM iterations to further improve the image quality. We set $\eta=0.02$ and initialize \{$\uv^k=\Tmat_{\Thetamat}(\ev)^{k-1}$, $\vv=0$\}, where $k$ is the iteration number that HSI deep denoising prior is first inserted.

\begin{figure}[htbp!]
\vspace{-3mm}
\centering
{\includegraphics[width=.96\linewidth]{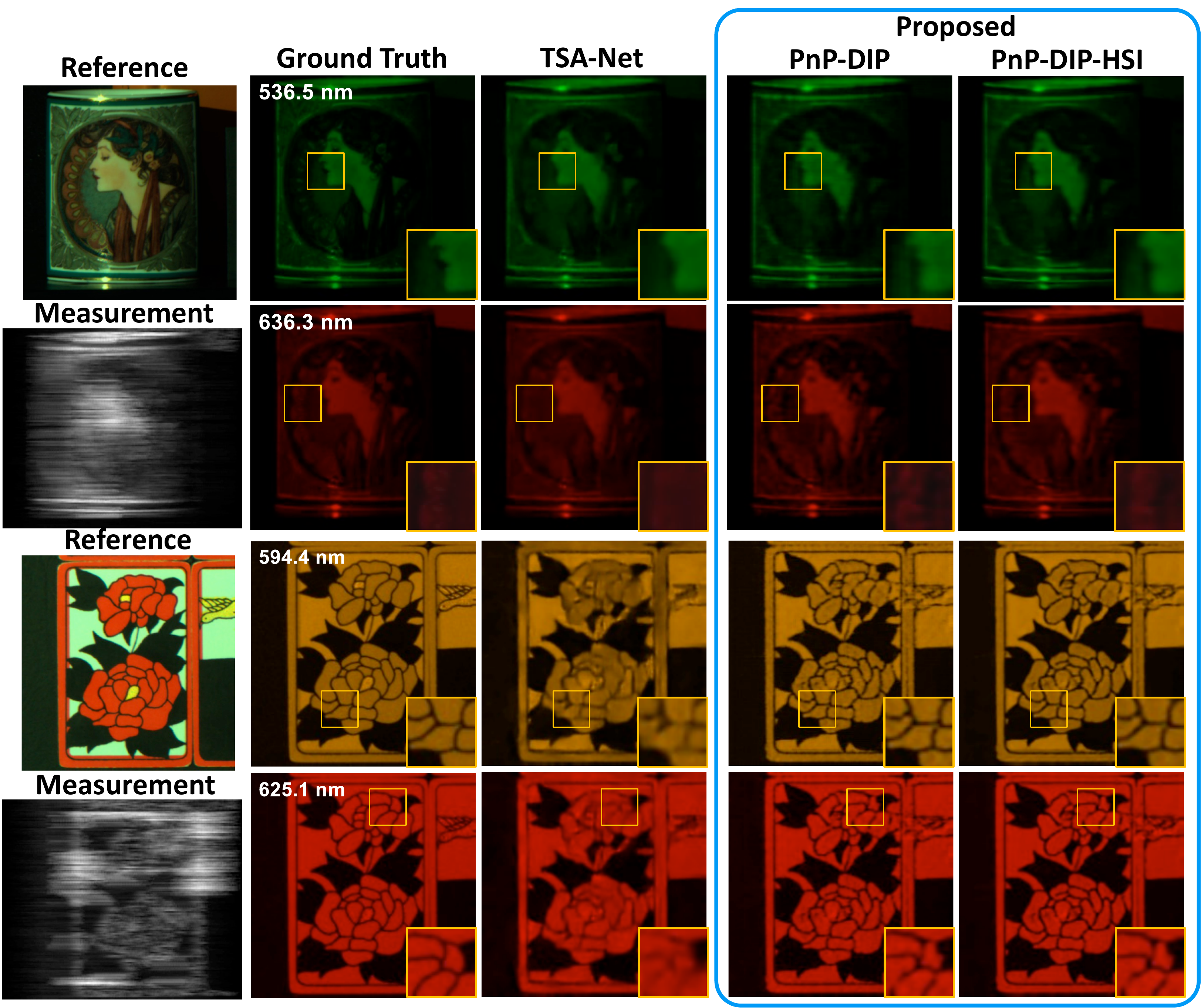}}
\vspace{-1mm}
\caption{Reconstructed {\bf synthetic data} ({\em Scene 1} and {\em 7}) with 2 spectral channels by 3 algorithms. Zoom in for better view.
}
\label{fig:simu_2}
\vspace{-5mm}
\end{figure}

\noindent \textbf{Datasets and Metric}
The testing datasets contain 10 scenes used in~\cite{Meng20ECCV_TSAnet} from KAIST~\cite{Choi17TOG}  with size $256\times256\times28$, \ie, 28 spectral bands with each one $256\times256$ pixels.
For the fair comparisons, we use the same {\bf real mask} as in~\cite{Meng20ECCV_TSAnet} to generate the measurements for recovering the synthetic data. 
There are two-pixel shifts between the neighboring spectral channels. 
Both Peak-Signal-to-Noise-Ratio (PSNR) and structural similarity (SSIM)~\cite{wang2004image} are employed to evaluate the quality of reconstructed spectral data-cube.

\noindent \textbf{Comparing Methods}
We compare our proposed methods (PnP-DIP and PnP-DIP-HSI) with other leading algorithms, including three optimization algorithms, \ie, TwIST~\cite{Bioucas-Dias2007TwIST}, ADMM-TV~\cite{Yuan16ICIP_GAP} and DeSCI~\cite{Liu18TPAMI}, a deep PnP method PnP-HSI~\cite{Zheng20_PRJ_PnP-CASSI}, and a state-of-the-art supervised method TSA-Net~\cite{Meng20ECCV_TSAnet}, in which a set of real data are reported\footnote{\url{https://github.com/mengziyi64/TSA-Net}} and we use them in the next section. 
We also compare with DeepRED~\cite{Mataev_2019_ICCVW_DeepRed}, which used both DIP and RED priors. We further compare with the auto-encoder approach proposed in~\cite{Choi17TOG}. However, due to different spectral channel numbers, we put the results in the supplementary materials (SM).

Table~\ref{Tab:simu} lists the PSNR and SSIM on the 10 scenes reconstructed by the aforementioned algorithms. It can be seen that the PSNR values of our proposed method PnP-DIP without using any training data are much higher than other optimization algorithms, DeepRED and the PnP-HSI recently developed in~\cite{Zheng20_PRJ_PnP-CASSI}. Even compared with the supervised method TSA-Net, the proposed PnP-DIP has a 0.33dB improvement in PSNR. However, the SSIM of PnP-DIP is lower than that of TSA-Net. By combining the pre-trained HSI deep denoiser~\cite{Zheng20_PRJ_PnP-CASSI} with DIP, the results of PnP-DIP-HSI show a further improvement, especially in SSIM, leading to the state-of-the-art results on both PSNR and SSIM.

We compare the spatial details and spectral accuracy of the above 8 algorithms on {\em Scene 9}, with results shown in Fig.~\ref{fig:simu_1}. The recovered spectral images are converted to synthetic-RGB (sRGB) via the CIE color matching function~\cite{smith1931cie}. It can be seen that the optimization algorithms suffer from the blurry on the horizontal axis, which might be caused by the shifting effects of the disperser in the system. PnP-HSI is unable to fully exert its advantage due to the less-than-perfect initialization of ADMM-TV. Compared with DeepRED and TSA-Net, the results of our PnP-DIP show sharper edges and better visual qualities. 
In addition, the reconstructed spectral curves of the proposed methods have a higher correlation with the reference spectra. Fig.~\ref{fig:simu_2} shows the comparisons of the proposed PnP-DIP, PnP-DIP-HSI and the TSA-Net on two other scenes. It can be observed that although TSA-Net can provide visually decent results, edge blurring and details loss appear in some regions. PnP-DIP can recover most of spatial details, but with some local noise. PnP-DIP-HSI is benefiting from both the DIP and the deep denoiser, and thus can mitigate both details loss and the noise effect.

\noindent \textbf{Running Time}
The projection step in the PnP framework can be updated
very efficiently, and the time consuming step is the updating of $\Thetamat$ by back-propagation. In our implementation, the average number of the DIP inner loop is about 900, and the outer loop is set to be 80 times, \ie, 80 ADMM iterations. In this case, the average running time is about 1 hour on a server with i7 CPU, 64 RAM and an Nvidia RTX3090 GPU. The running time can be saved by running it in parallel and initializing the result by ADMM-TV.
Compared with other unsupervised (model-based) algorithms, our proposed approach is much faster than DeSCI (which needs more than 3 hours) and provides better results.
When a real-time reconstruction is desired, supervised deep networks after training might be the right choice.
\begin{table}[htbp!]
\vspace{-2mm}
\caption{Average PSNR and SSIM of sole DIP, PnP-DIP with single fidelity term in~\eqref{eq:xTheta_joint1} and the proposed PnP-DIP with double fidelity terms in~\eqref{eq:xTheta_joint3}.}
\vspace{-3mm}
\begin{center}
\resizebox{.47\textwidth}{!}{
\begin{tabular}{l|c|c|c}
\hline
Approach &  Sole DIP & 
\shortstack{PnP-DIP\\ (Single Fidelity)} & \shortstack{PnP-DIP\\ (Double Fidelity)}\\
\hline
PSNR/SSIM & 26.99, 0.777 & 28.87, 0.824 &  30.48, 0.854\\
\hline
\end{tabular}
}
\end{center}
\vspace{-8mm}
\label{Tab:SFvsDF}
\end{table}

\subsection{Ablation Study}
In this section, we perform a comprehensive comparison and ablation study using different modules and configurations in our proposed algorithm.

\noindent \textbf{DIP vs. Deep Decoder}
Firstly, we investigate the network structure of $\Tmat_{\Thetamat}(\ev)$.
We compare the performance of DIP with the deep decoder~\cite{heckel_deep_2018}, which are two well-known untrained neural network for image restoration, with the other parameters keeping the same in the framework. The average PSNR and SSIM of PnP-DIP (30.48dB, 0.854) are better than the PnP framework using deep decoder (28.44dB, 0.819). Please refer to more detailed comparisons in the SM. We analyze the reason of the performance gap and this might be due to the following two factors.  $i$) Deep decoder is originally designed for image compression, thus containing much less network parameters compared with the U-net in DIP. $ii$) Different from the inverse problems solved by deep decoder in~\cite{heckel_deep_2018}, which only recover a 2D or RGB image, our task aims to recover a 3D spectral cube. Therefore, more layers and parameters are needed in the deep prior network.

\begin{table}[htbp!]
\vspace{-2mm}
\caption{Average PSNR and SSIM of 3 methods on the 10 synthetic data with different {\em Poisson} noise levels.}
\vspace{-3mm}
\begin{center}
\resizebox{.45\textwidth}{!}{
\begin{tabular}{l|c|c|c}
\hline
Noise level &  TSA-Net & PnP-DIP & PnP-DIP-HSI\\
\hline
No noise & 30.15, 0.893 & 30.48, 0.854 & 31.30, 0.901\\
\hline
SNR=30dB & 27.38, 0.801 & 28.91, 0.783 & 29.71, 0.860\\
\hline
SNR=25dB & 24.15, 0.711 & 27.73, 0.731 & 28.69, 0.838\\
\hline
\end{tabular}
}
\end{center}
\vspace{-6mm}
\label{Tab:Noise}
\end{table}

\noindent \textbf{Single Fidelity vs. Dual Fidelity in DIP}
Hereby, we show the results of DIP using single fidelity term in Eq.~\eqref{eq:xTheta_joint1} (the derivation of the solution is shown in the SM) and the proposed dual fidelity terms ($\|\yv-\Hmat\xv\|_2^2$ and $\|\yv-\Hmat\Tmat_{\Thetamat}(\ev)\|_2^2$) in Eq.~\eqref{eq:xTheta_joint3}.
Meanwhile, we also compare with the directly using DIP for spectral SCI reconstruction. The results are summarized in Table \ref{Tab:SFvsDF}. %
It can be seen that the sole DIP can only achieve an average PSNR at 26.99dB, which is 3.5dB less than our proposed method. In addition, the single fidelity result is 1.6dB lower than our proposed model on PSNR.

\noindent \textbf{Noise Robustness}
We further test the noise robustness of the proposed methods by recovering the images from the measurements contaminated by {\em Poisson noise} with different signal-to-noise ratio (SNR). It is well known that Poisson noise is a better noise model in real systems. Table \ref{Tab:Noise} compares the results of the proposed methods with TSA-Net under the SNR of 30dB and 25dB. It can be seen that the performance degradation of our methods is less than TSA-Net (\ie, 2.75dB and 2.61dB for PnP-DIP and PnP-DIP-HSI, respectively, and 6.00dB for TSA-Net at SNR=25dB). Therefore our methods are more robust to noise than the supervised methods.


We also noticed that different up/downsampling operations in U-net used in DIP will affect the results and more ablation studies are presented in the SM as well as comparisons of different priors.

\begin{figure*}[htbp!]
\centering
{\includegraphics[width=1\linewidth]{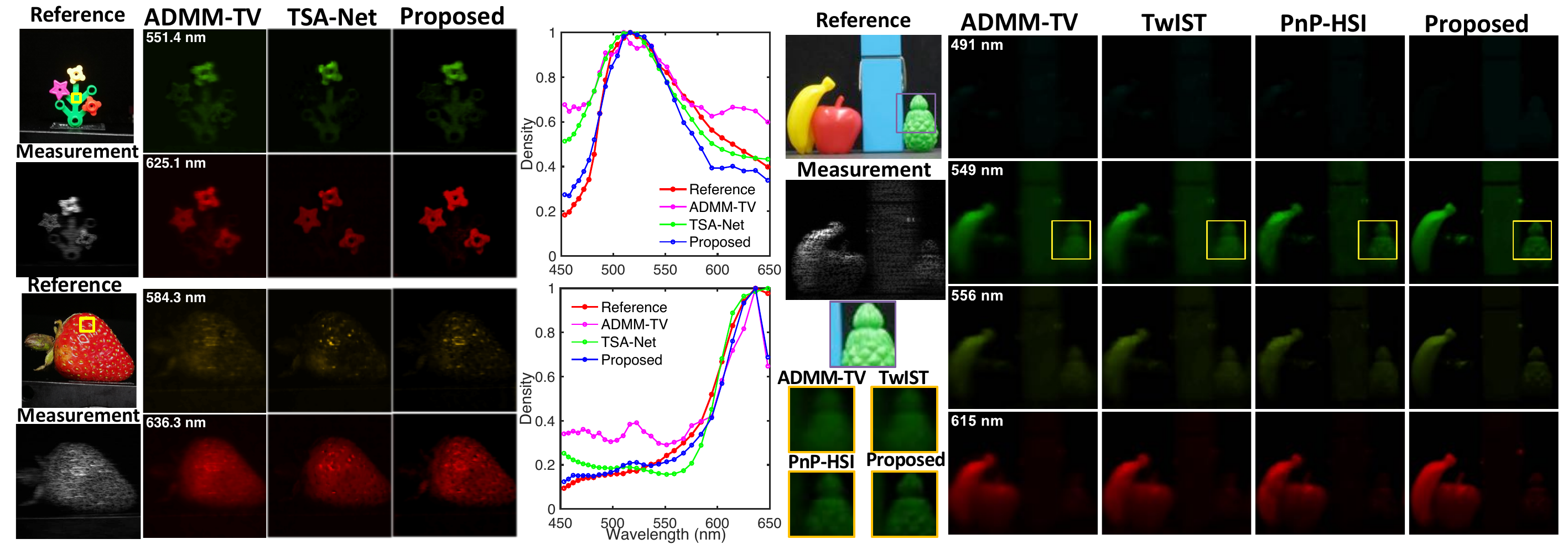}}
\caption{Reconstructed {\bf real data} of CASSI datasets by different algorithms. Left: {\em Lego plant} and {\em Strawberry} with 2 out of 28 spectral channels and spectral curves of the selected regions. Right: {\em Object} with 4 out of 33 spectral channels. Zoom in for better view.
}
\label{fig:real_2}
\vspace{-5mm}
\end{figure*}
\begin{figure}[htbp!]
\centering
{\includegraphics[width=1\linewidth]{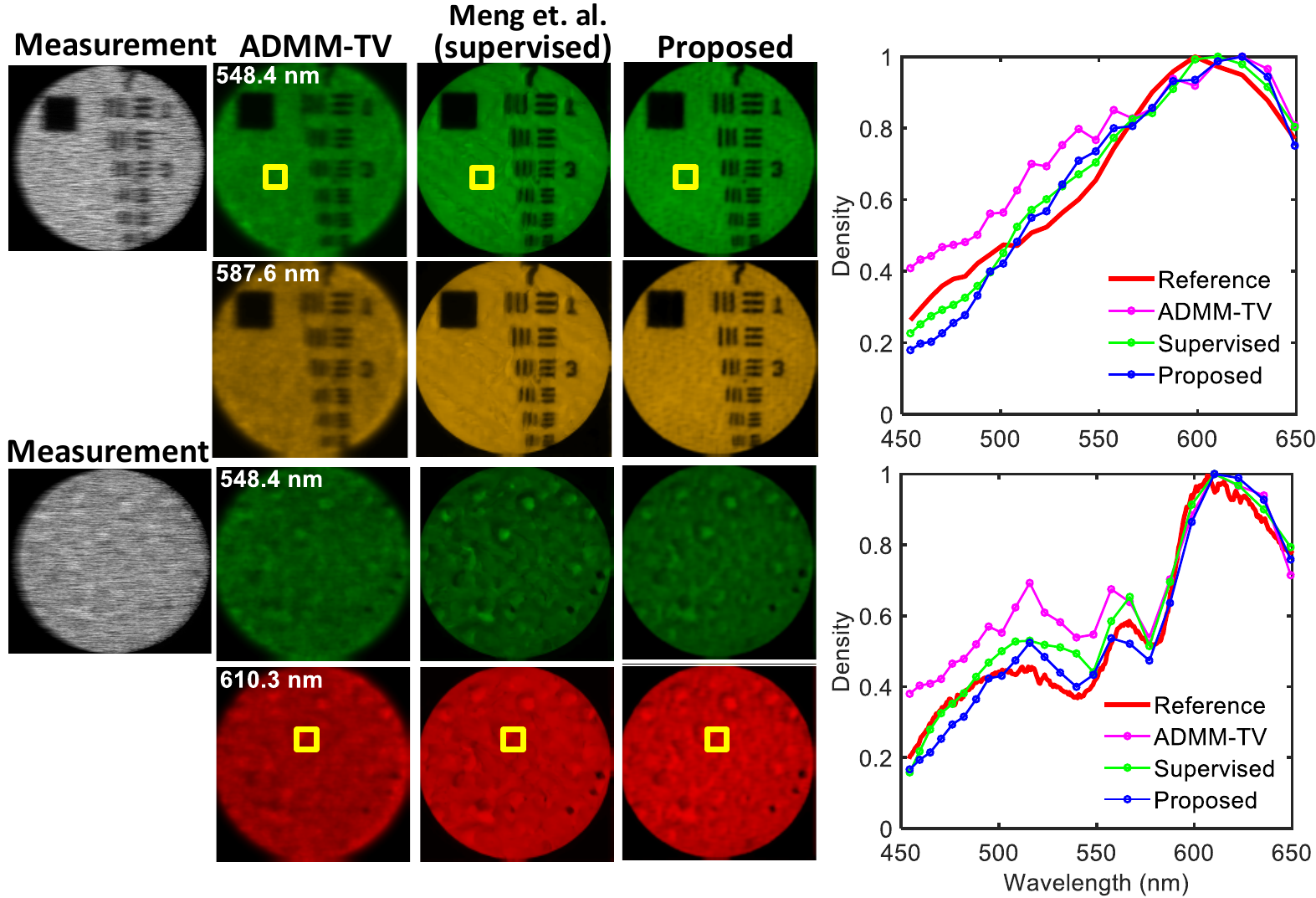}}
\caption{Reconstructed {\bf real data} of 2 scenes, {\em Resolution target} and {\em Red blood cell}, with 2 out of 24 spectral channels and spectral curves of the selected regions by 3 algorithms.
}
\label{fig:real_3}
\vspace{-6mm}
\end{figure}


\subsection{Real Data Results \label{Sec:real}}
We apply our proposed methods on the real data captured by three spectral SCI systems, \ie, the most recently built CASSI system~\cite{Meng20ECCV_TSAnet}, the original CASSI system~\cite{kittle2010multiframe} and the compressive multispectral endomicroscope~\cite{meng2020snapshot}\footnote{\url{https://github.com/mengziyi64/SMEM}}. 
Note that though the underlying principle of these systems is the same, it is challenging to find training data of endomicoscopy. 
Thereby, our proposed self-supervised algorithm is a perfect fit for this kind of tasks. 
Considering the large scale reconstruction of the real data and the measurement noise, we reset the process and parameters of our algorithm. Specifically, we firstly use TV prior in the PnP framework to obtain a result, which serves as a warm starting point for PnP-DIP and saves the running time.

\noindent \textbf{CASSI Data Set}
We first show the results on the datasets captured by the CASSI system in~\cite{Meng20ECCV_TSAnet}. The recovered spectral cube contains 28 spectral channels with the size of $550\times550$. We compare the proposed method with ADMM-TV and TSA-Net on {\em Strawberry} and {\em Lego plant} datasets, as shown in the left of Fig.~\ref{fig:real_2}.
It can be observed that compared with the TSA-Net, our reconstruction has less artifacts. In addition, we show the recovered spectral curves corresponding to the selected regions. Our proposed method provides a higher spectral accuracy compared with the other two algorithms. The upper part of Fig.~\ref{fig:real_1} shows another reconstructed scene {\em Plant} with 5 spectral channels. We can see that the results of the proposed self-supervised method are visually better than the results of TSA-Net.
We show the real data ({\em Object}) with the size of $210\times256\times33$ captured by the original CASSI system~\cite{kittle2010multiframe} in the right of Fig.~\ref{fig:real_2}, where again we can see our method recovers better spatial details compared with ADMM-TV, TwIST and PnP-HSI.
%


\noindent \textbf{Endomicroscopy Data}
Lastly, we apply the proposed method on the endomicroscopy data~\cite{meng2020snapshot}. This data was captured by a compressive multispectral endomicroscopy system, which obtains images by a fiber bundle and a spectral SCI system. The captured measurements are used to reconstruct the multispectral endoscopic images with the size of $660\times660\times24$.
We compare the proposed method (with using HSI prior) with ADMM-TV and a trained deep neural network (DNN)~\cite{meng2020snapshot} on two data, \ie, {\em Resolution target} and {\em Red blood cell}, as shown in Fig.~\ref{fig:real_3}. It can be observed that our reconstructed images achieve higher spatial resolution and cleaner details. The reconstructed spectra of our method are more accurate compared with other algorithms. 
Additionally, the lower part of Fig.~\ref{fig:real_1} shows another reconstructed scene {\em Dog olfactory membrane section} with 5 spectral channels, where we can see the results of the proposed supervised method has less artifacts.


\section{Conclusions \label{Sec:con}}
We have proposed a self-supervised algorithm for the reconstruction of spectral snapshot compressive imaging. The proposed framework uses an untrained neural network to learn a prior directly from the compressed measurement captured by the snapshot compressive imaging system. Therefore, the proposed algorithm does not need any training data.
We integrate this untrained deep network based prior into the plug-and-play framework and solve it by the alternating direction method of multipliers algorithm.
By using a different formulation from existing algorithms, we have achieved competitive results to those of supervised deep learning based algorithms, which need extensive training data.
Furthermore, we have incorporated the proposed framework with a recently pre-trained hyperspectral images deep denoising network to achieve a joint reconstruction regime. 
This joint algorithm has provided state-of-the-art results on both synthetic and real datasets from  different spectral snapshot compressive imaging systems.

Regarding the future work, we believe that our proposed self-supervised framework can also be extended to the video SCI reconstruction~\cite{Patrick13OE,Yuan14CVPR,Qiao2020_APLP,Cheng2021_CVPR_ReverSCI,Cheng20ECCV_Birnat,Wang2021_CVPR_MetaSCI,Zheng2021_Patterns}.


{\small
\bibliographystyle{ieee_fullname}

}

\section{Derivation of Single Fidelity Formulation}

Derivation of the Solution for Single Fidelity Formulation in Eq. (8) in the main paper:
ADMM solves the (8) by splitting it into the following subproblems:
   \begin{list}{\labelitemi}{\leftmargin=8pt \topsep=0pt \parsep=0pt}
    \item Given $\Thetamat$ and $\bv$, $\xv$ is solved by 
    \begin{equation}
       \hat\xv = \argmin_{\xv} \|\xv - \Tmat_{\Thetamat}(\ev) -\bv\|_2^2  + \frac{\lambda}{\mu} R(\xv).
    \end{equation}
    This is a traditional denoising problem and can be solved by the PnP algorithm given the prior $R(\xv)$, \ie,
    \begin{equation}
        \hat\xv  = {\cal D}_{\sigma} (\Tmat_{\Thetamat}(\ev) +\bv).
    \end{equation}
    where ${\cal D}_{\sigma}$ denotes the denoising operator being used and $\sigma$ is the estimated noise level depending on $\lambda/\mu$.
    \item Given $\xv$ and $\bv$, optimizing $\Thetamat$ leads to the following problem:
    \begin{equation}
        \hat\Thetamat = \argmin_{\Thetamat} \frac{1}{2}\|\yv - \Hmat\Tmat_{\Thetamat}(\ev)\|_2^2 +\mu \|\xv - \Tmat_{\Thetamat}(\ev) -\bv\|_2^2, \label{eq:opt_theta}
    \end{equation}
which can be solved by the  back-propagation optimization as in DIP, modified by a proximity regularization that forces $\Tmat_{\Thetamat}(\ev)$ to be close to $\xv-\bv$. 
For the U-net being used in our implementation, instead of only minimizing the first term in \eqref{eq:opt_theta} as in the loss function, we used both terms as the loss function.
This learned $\Tmat_{\Thetamat}(\ev)$ is thus playing the role of: i) denoising $\xv-\bv$, and ii) minimizing the measurement loss $\yv - \Hmat \Tmat_{\Thetamat}(\ev)$. 
\item Optimizing $\bv$ is given by
\begin{equation}
     \bv^{k+1} =  \bv^k -(\xv^k - \Tmat_{\Thetamat^k}(\ev)),
\end{equation}
where the superscript $k$ denotes the iteration number. 
\end{list}
Note that these three steps are performed iteratively and each of them can have their own inner loops such as the $\Thetamat$ optimization. 

\section{Ablation Study Results}

\subsection{DIP vs. Deep Decoder}
\begin{figure}[!htbp]
\centering
\renewcommand\thefigure{M1}
{\includegraphics[width=.98\linewidth]{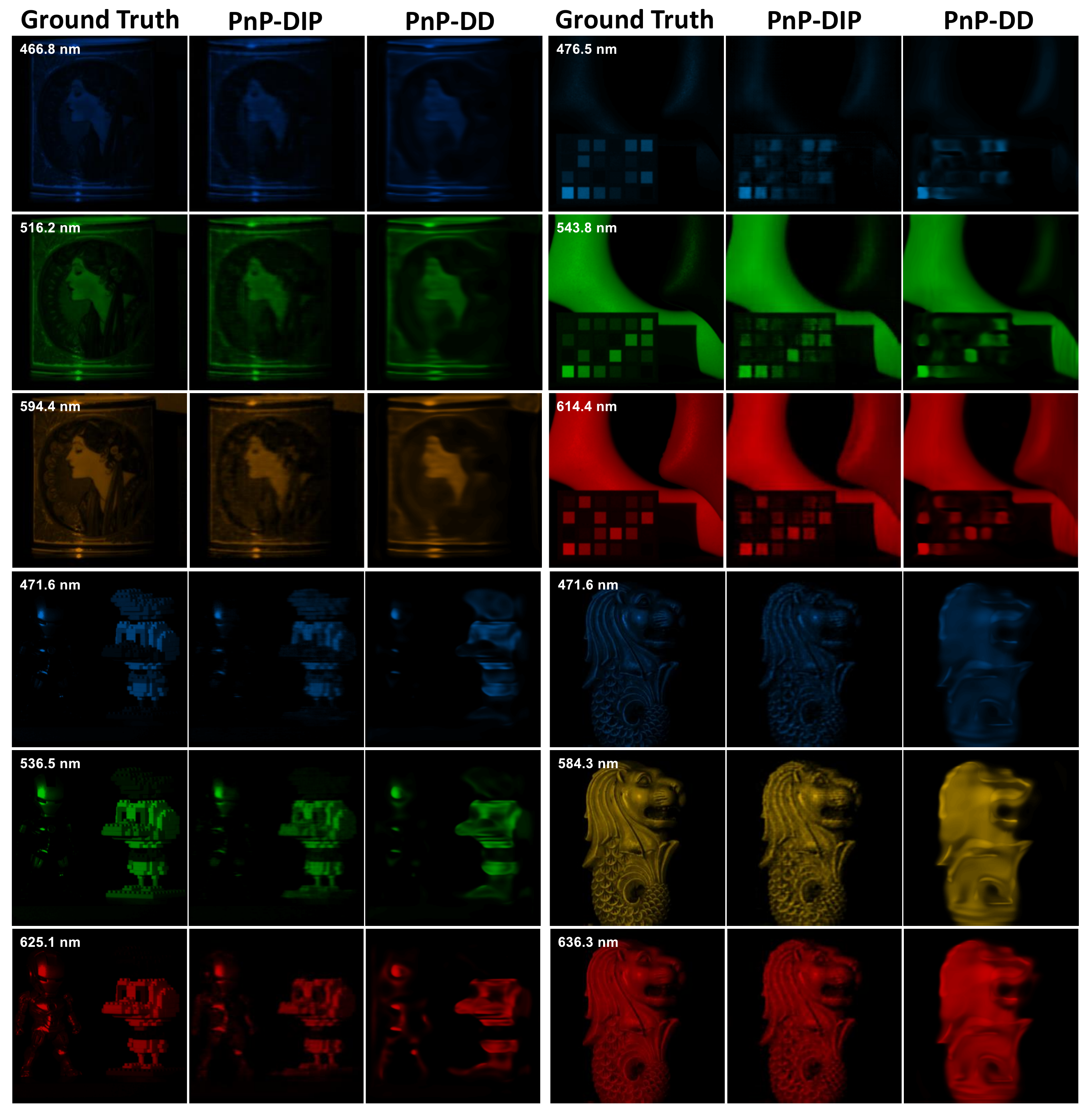}}
\caption{Reconstructed results of 4 synthetic data with 3 spectral channels by the proposed self-supervised methods using DIP and deep decoder as the prior, respectively. }
\label{fig:sm1}
\vspace{-3mm}
\end{figure}

We visualize the results of the proposed self-supervised methods using DIP~\cite{Ulyanov_2018_CVPR} and deep decoder (DD)~\cite{heckel_deep_2018} as the prior (PnP-DIP and PnP-DD), as shown in Fig.~\ref{fig:sm1}. It can be seen seen that PnP-DD provide a good reconstruction on some smooth regions of the images, but for the regions with many spatial details, there are significant artifacts and over-smoothness. As mentioned in the main paper, this might be caused by the lack of the network parameters of the deep decoder.

\begin{figure*}[!htbp]
\centering
\renewcommand\thefigure{M2}
{\includegraphics[width=1\linewidth]{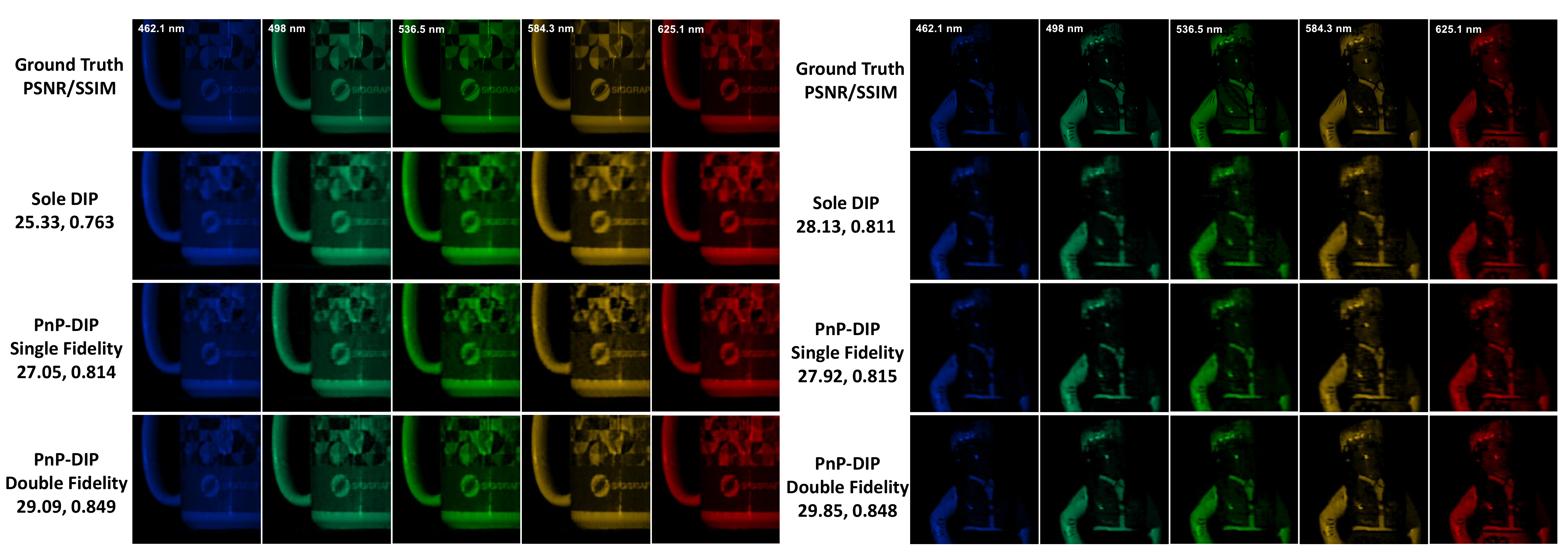}}
\vspace{-4mm}
\caption{Reconstructed results of two synthetic data with 5 spectral channels by the sole DIP, PnP-DIP with single fidelity term and the proposed PnP-DIP with double fidelity terms.}
\label{fig:sm2}
\vspace{-4mm}
\end{figure*}

\subsection{Single Fidelity vs. Dual Fidelity in DIP}
Fig.~\ref{fig:sm2} compare the results of the sole DIP (the directly using DIP), PnP-DIP with single fidelity term and the proposed PnP-DIP with double fidelity terms. It can be seen that the reconstructed results of the proposed PnP-DIP with double fidelity terms have clearer detials, as well as less noise and artifacts.

\subsection{Incorporating DIP with TV Prior}
Without considering pre-trained HSI denoiser, the previous self-supervised results are obtained by only using DIP as the prior in our proposed PnP framework (PnP-DIP). Here we incorporate the widely used TV prior with DIP to form a joint PnP framework, namely PnP-DIP-TV. We initial the parameters by \{$\uv=\Hmat\ts\yv$, $\vv=0$, $\eta=0.01$\}, and other parameters keep the same as before. We reduce the effect of TV prior gradually as the increasing of iterations by scaling $\eta$ by 0.95 each ADMM iteration. Finally, the average results of PnP-DIP-TV (30.44dB, 0.852) is very close to PnP-DIP (30.48dB, 0.854). 
This gives us the following observations;
$i$) Though TV is widely used and can achieve good results in most tasks, DIP is powerful to learn a stronger prior. Similar case will happen for other pre-trained priors such as sparsity.
$ii$) Even at the first few iterations, TV will help the reconstruction, the final results will rely on DIP. 
Therefore, we recommend that in our spectral SCI reconstruction, PnP-DIP can be used as a new baseline without any training data.
However, we do notice that in real data experiments, TV will help the reconstruction, which may be due to the measurement noise. 

\subsection{Choice of the Up/downsampling}
U-net is an encoder-decoder scenario, and thus up/downsampling is playing a pivotal role.  
As the crucial components of U-net, pooling and upsampling change the scale and depth of the feature maps. Upsampling is usually implemented by unlearned forms, (such as bilinear and PixelShuffle~\cite{shi2016real}) and learned convolutional filters (transposed convolution or ConvTranspose~\cite{zeiler2014visualizing}). We compare the results using different upsampling in the DIP network, shown in the upper part Table~\ref{Tab:upsampling}. It can be seen that the network using ConvTranspose achieves the highest performance. For the downsampling, we find that the average-pooling provide a better results compared with max-pooling with comparison shown in the lower part in Table~\ref{Tab:upsampling}.
Therefore,  ConvTranspose and average-pooling are used in our experiments.

\begin{table}
\renewcommand\thetable{M1}
\caption{Average PSNR and SSIM of the DIP network using different up/downsampling.}
\vspace{-2mm}
\begin{center}
\resizebox{.47\textwidth}{!}{
\begin{tabular}{l|c|c|c}
\hline
Upsampling & Conv2DTranspose & Bilinear & PixelShuffle\\
\hline\hline
PSNR/SSIM & 30.48, 0.854 & 29.69, 0.821 & 27.59, 0.824\\
\hline
Downsampling & Average-pooling & Max-pooling  \\
\hline
PSNR/SSIM & 30.48, 0.854 & 30.26, 0.848 \\
\hline
\hline
\end{tabular}
}
\end{center}
\vspace{-5mm}
\label{Tab:upsampling}
\end{table}

\begin{table*}[htbp!]
\renewcommand\thetable{M2}
\caption{Average PSNR and SSIM of ADMM-TV, PnP-HSI and PnP-DIP on two datasets used in~\cite{Zheng20_PRJ_PnP-CASSI} under two different simulation settings.}
\centering
\resizebox{1.35\columnwidth}{!}
{
\begin{tabular}{c|c|c|c|c}
\hline
Dataset &  Simulation setting & 
ADMM-TV & PnP-HSI & PnP-DIP\\
\hline
ICVL & Binary mask, 30-pixel shift & 32.56, 0.899 & 39.43, {\bf 0.974} & {\bf 40.72}, 0.970\\
\cline{2-5}
 & Real mask, 60-pixel shift & 29.01, 0.867 & 32.91, 0.930 & {\bf 37.27}, {\bf 0.954}\\
\hline
KAIST & Binary mask, 30-pixel shift & 37.25, 0.957 & 39.15, {\bf 0.974} & {\bf 41.79}, {\bf 0.974}\\
\cline{2-5}
 & Real mask, 60-pixel shift & 34.25, 0.941 & 34.92, 0.954 & {\bf 38.74}, {\bf 0.962}\\
 \hline
\end{tabular}}
\label{Tab:t2}
\end{table*}

\section{Supplementary Results}

\subsection{PnP-DIP vs. PnP-HSI}

As shown in Table 1 in the main paper, the average result of PnP-HSI~\cite{Zheng20_PRJ_PnP-CASSI} ([42] in the main paper) has an about 5dB gap with the proposed PnP-DIP.
The main reason causing the less-than-perfect results of PnP-HSI is the simulation setting. We used the {\em real captured mask} and {\bf a larger mask-shift range (54 pixels)}. PnP-HSI is heavily dependent on the initialization results of ADMM-TV, which is not good in our simulation setting. The results of PnP-HSI usually cannot converge well (generating artifacts) when using a bad initialization. This is why we use HSI denoiser in only the last few ADMM iterations. Our simulation setting is closer to the real systems compared with~\cite{Meng20ECCV_TSAnet}, and our results indicate that PnP-HSI is getting degraded when the shifting pixels are larger.

\begin{figure}[!htbp]
\centering
\renewcommand\thefigure{M3}
\vspace{-2mm}
\includegraphics[width=.98\linewidth]{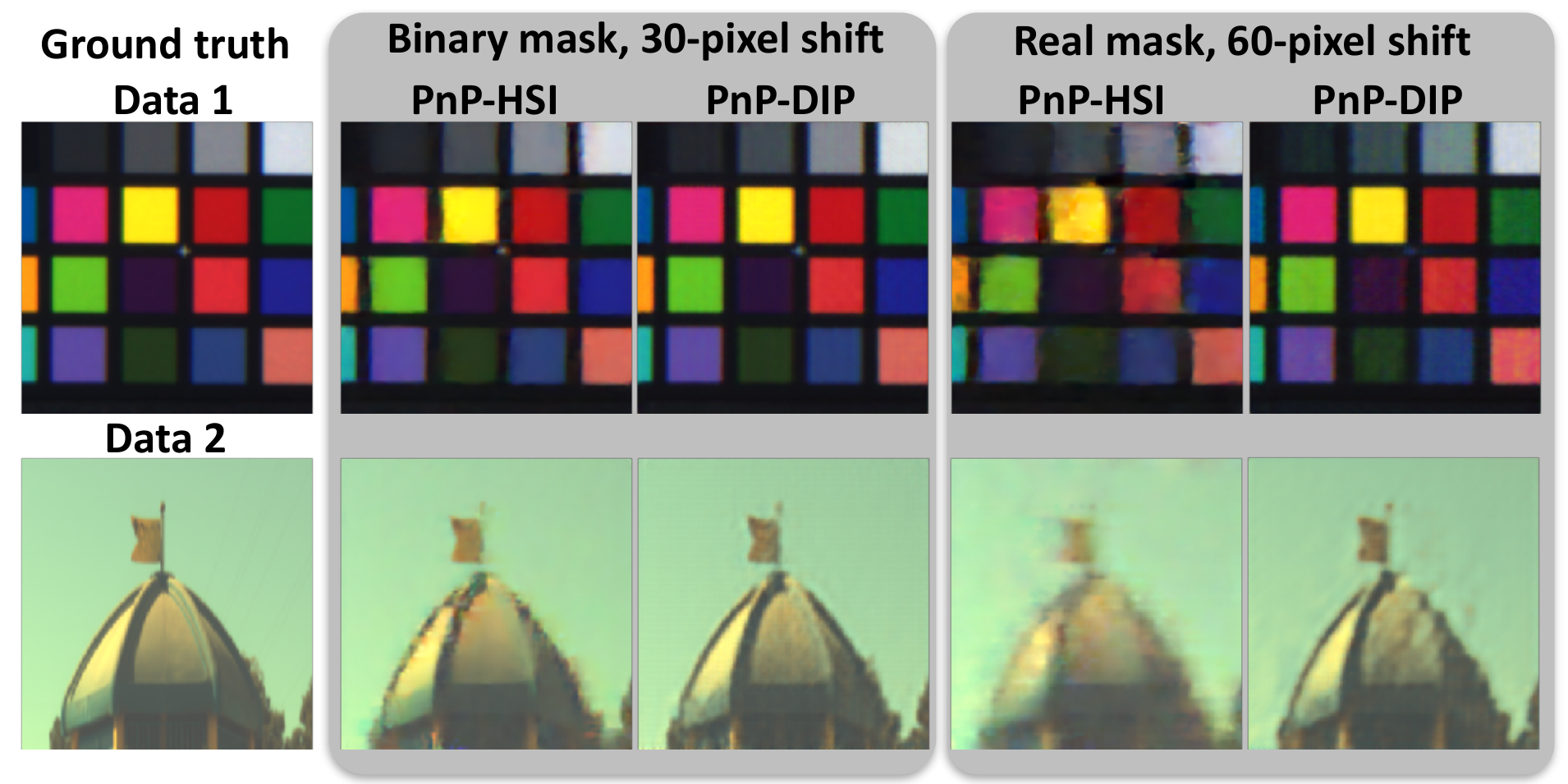}
\caption{Result comparison of PnP-HSI and PnP-DIP on two data from~\cite{Zheng20_PRJ_PnP-CASSI} under two different simulation settings.}
\label{fig:sm_R1}
\vspace{-3mm}
\end{figure}

For verifying the analysis, we give the results of the datasets used by~\cite{Zheng20_PRJ_PnP-CASSI} in Table \ref{Tab:t2} and Fig.~\ref{fig:sm_R1}.
Specifically, when the mask shifting is small, PnP-DIP and PnP-HSI are providing similar results, but when the mask shifting is big, our proposed PnP-DIP outperforms PnP-HSI by 4.36dB and 3.82dB, respectively on the two datasets, respectively. 

\begin{figure}[!htbp]
\centering
\renewcommand\thefigure{M4}
\includegraphics[width=.98\linewidth]{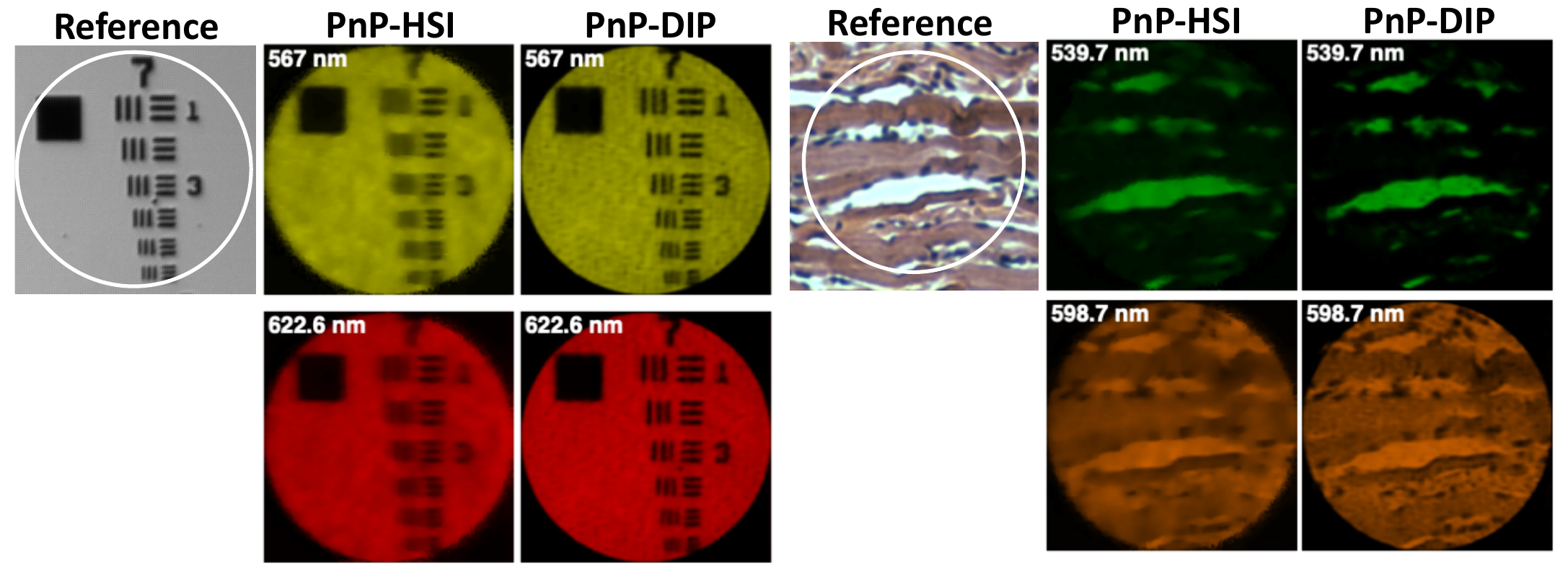}
\caption{Result comparison of PnP-HSI and PnP-DIP on endomicroscapy data in\cite{meng2020snapshot}.}
\label{fig:sm_R2}
\vspace{-5mm}
\end{figure}

This has also been verified by the real data results (Fig.~5 in the main paper and Fig.~\ref{fig:sm_R2} in this SM) where a large mask shifting was used.

\subsection{PnP-DIP vs. Autoencoder}
\begin{figure}[!htbp]
\centering
\renewcommand\thefigure{M5}
\includegraphics[width=.98\linewidth]{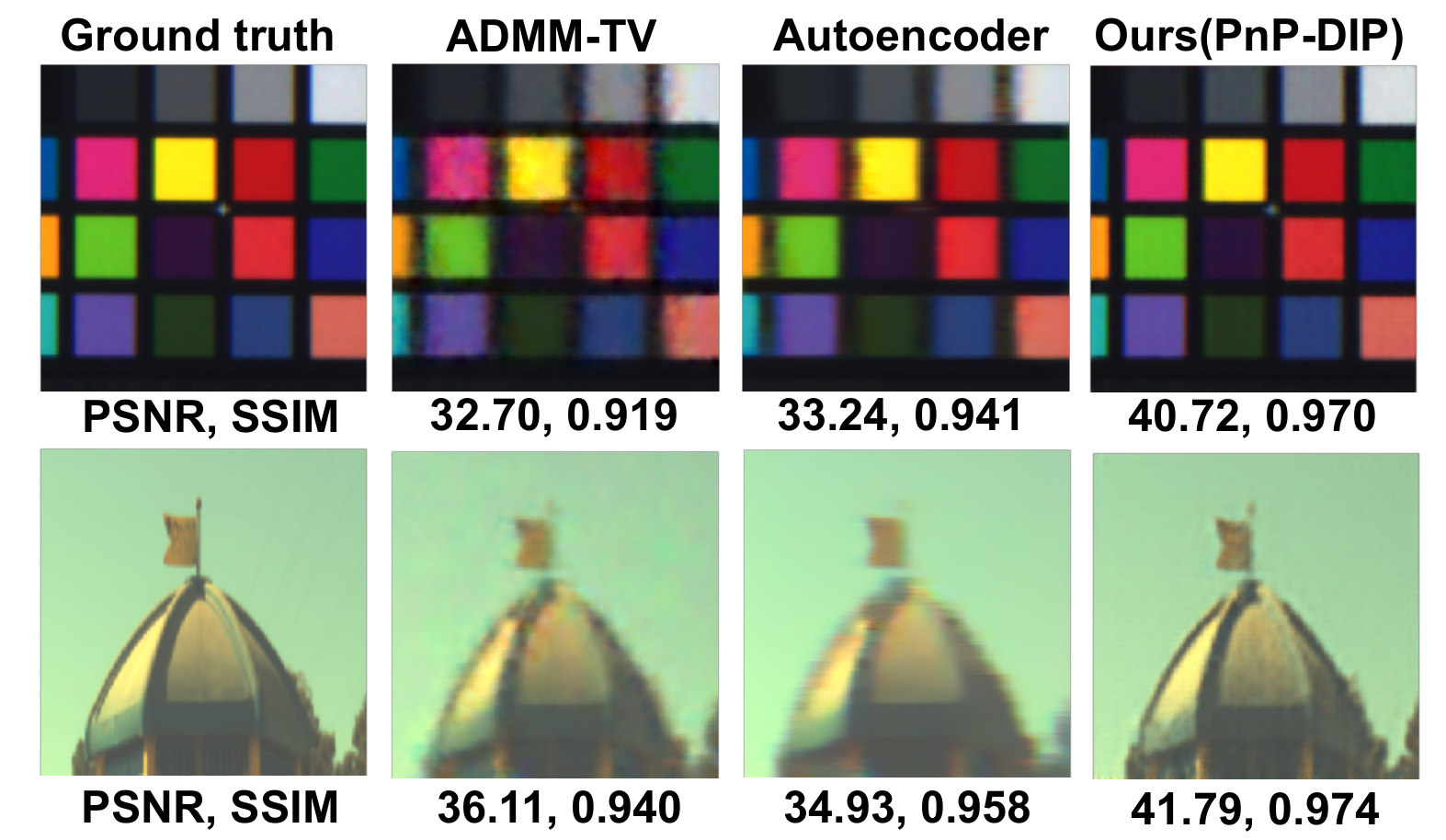}
\caption{Comparison (sRGB) of ADMM-TV [37], Autoencoder [6] and our PnP-DIP on two datasets used in~\cite{Zheng20_PRJ_PnP-CASSI}.}
\label{fig:R2}
\vspace{-3mm}
\end{figure}

We compare our proposed PnP-DIP with Autoencoder~\cite{Choi17TOG} on the datasets used in~\cite{Zheng20_PRJ_PnP-CASSI}. We use binary mask in simulation, and the shift range is 30-pixel. Fig.~\ref{fig:R2} shows the sRGB results of ADMM-TV~\cite{Yuan16ICIP_GAP}, Autoencoder~\cite{Choi17TOG} and our PnP-DIP. It can be seen that our method achieves much better results. Autoencoder suffers from the spatial blur in this single-disperser CASSI model, which is different from the dual-disperser CASSI model mainly used in~\cite{Choi17TOG}. We will put the results into the final paper.

\subsection{Results on the Synthetic Data}

\begin{figure*}[htbp!]
\centering
\renewcommand\thefigure{M6}
{\includegraphics[width=.98\linewidth]{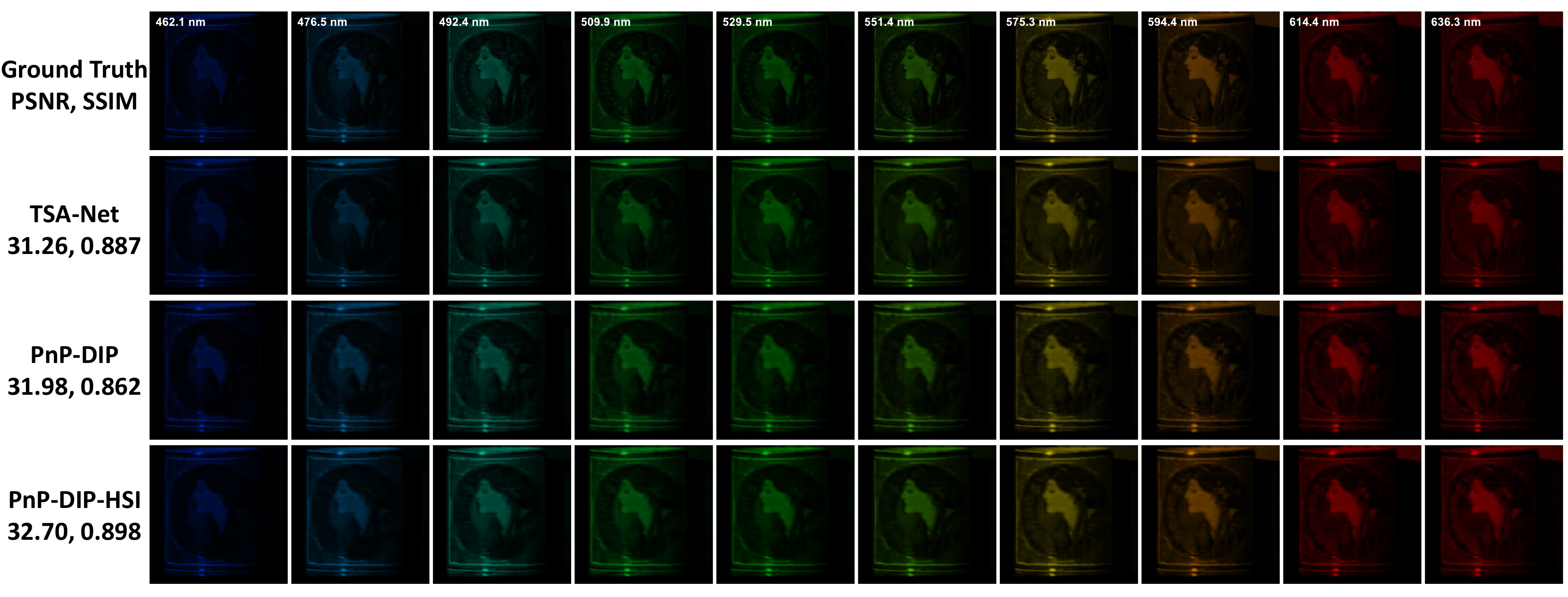}}
\caption{The results of the synthetic data {\em Scene 1} with 10 spectral channels reconstructed by TSA-Net and the proposed PnP-DIP and PnP-DIP-HSI.}
\label{fig:sm3}
\end{figure*}

\begin{figure*}[htbp!]
\centering
\renewcommand\thefigure{M7}
{\includegraphics[width=.98\linewidth]{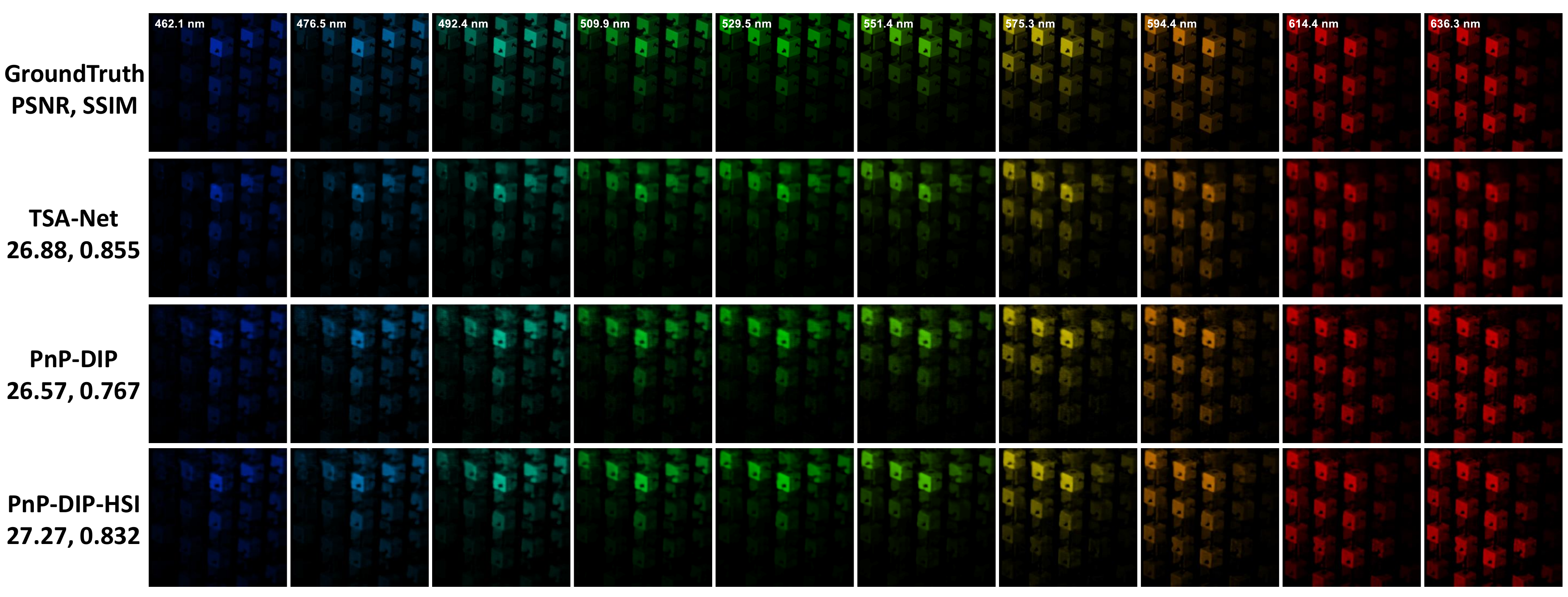}}
\caption{The results of the synthetic data {\em Scene 2} with 10 spectral channels reconstructed by TSA-Net and the proposed PnP-DIP and PnP-DIP-HSI.}
\label{fig:sm4}
\end{figure*}

\begin{figure*}[htbp!]
\centering
\renewcommand\thefigure{M8}
{\includegraphics[width=.98\linewidth]{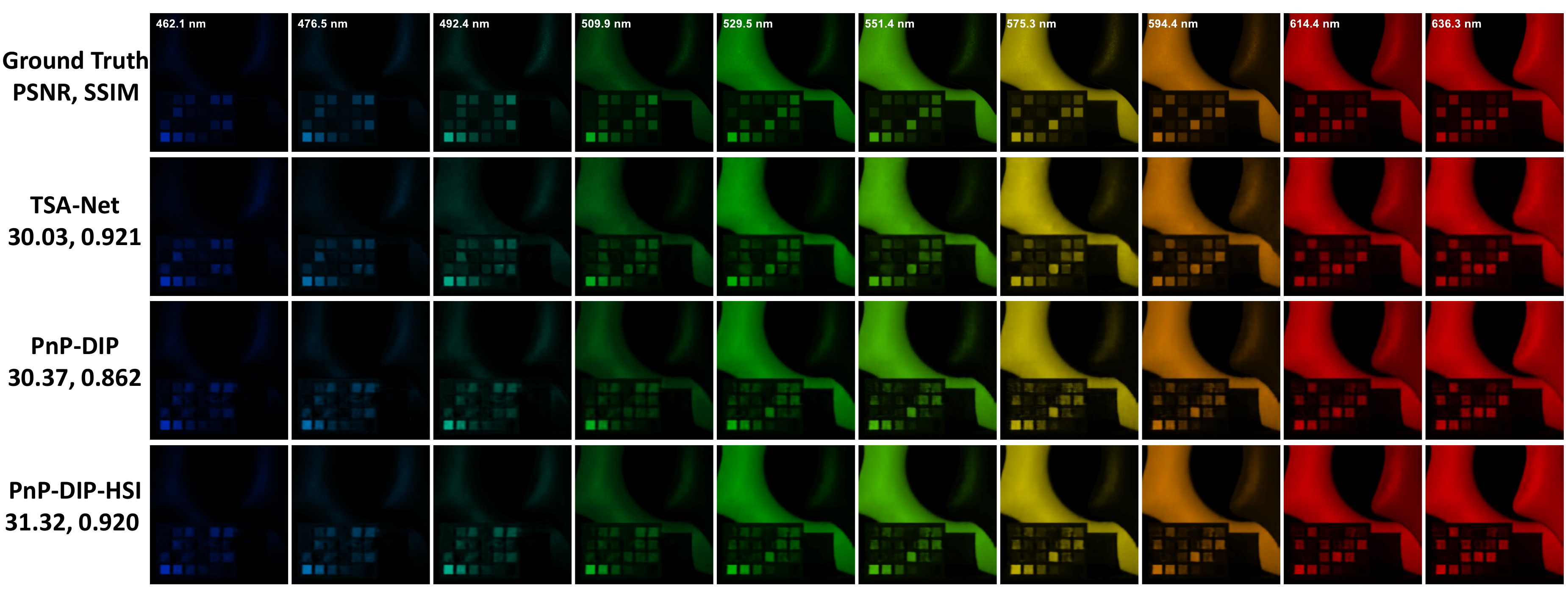}}
\caption{The results of the synthetic data {\em Scene 3} with 10 spectral channels reconstructed by TSA-Net and the proposed PnP-DIP and PnP-DIP-HSI.}
\label{fig:sm5}
\end{figure*}

\begin{figure*}[htbp!]
\centering
\renewcommand\thefigure{M9}
{\includegraphics[width=.98\linewidth]{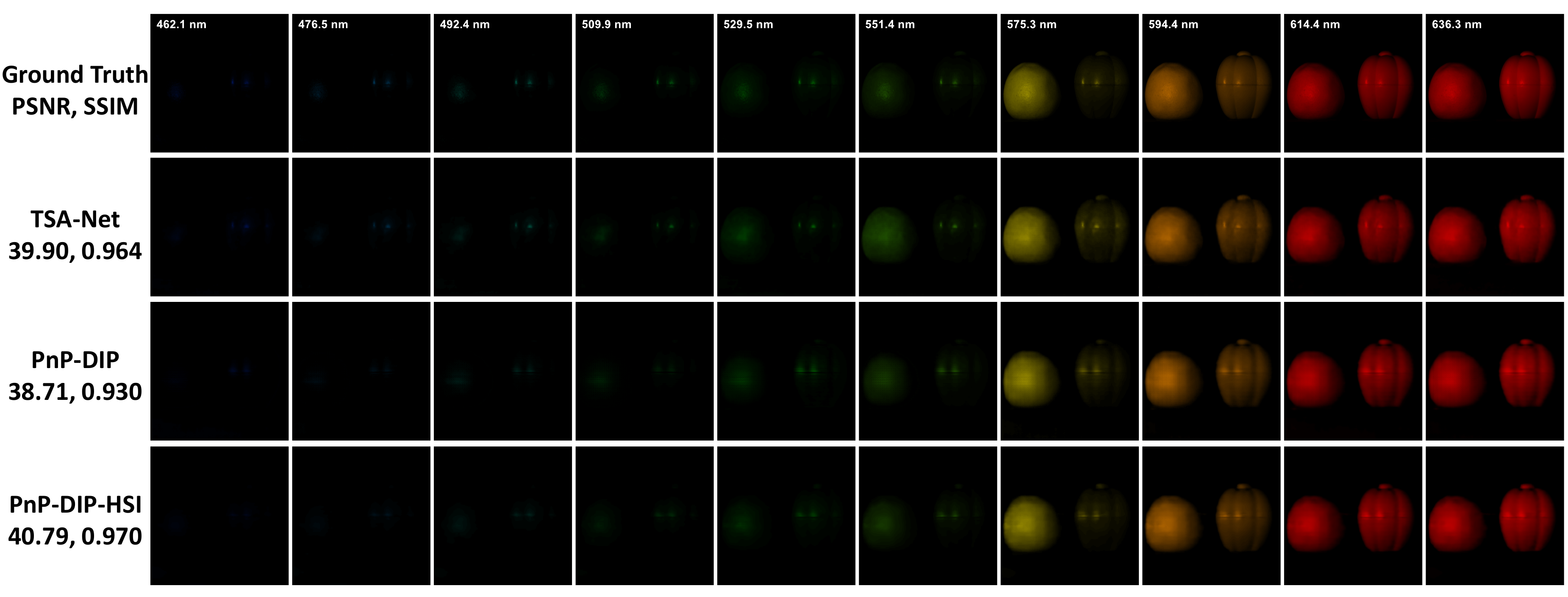}}
\caption{The results of the synthetic data {\em Scene 4} with 10 spectral channels reconstructed by TSA-Net and the proposed PnP-DIP and PnP-DIP-HSI.}
\label{fig:sm6}
\end{figure*}

\begin{figure*}[htbp!]
\centering
\renewcommand\thefigure{M10}
{\includegraphics[width=.98\linewidth]{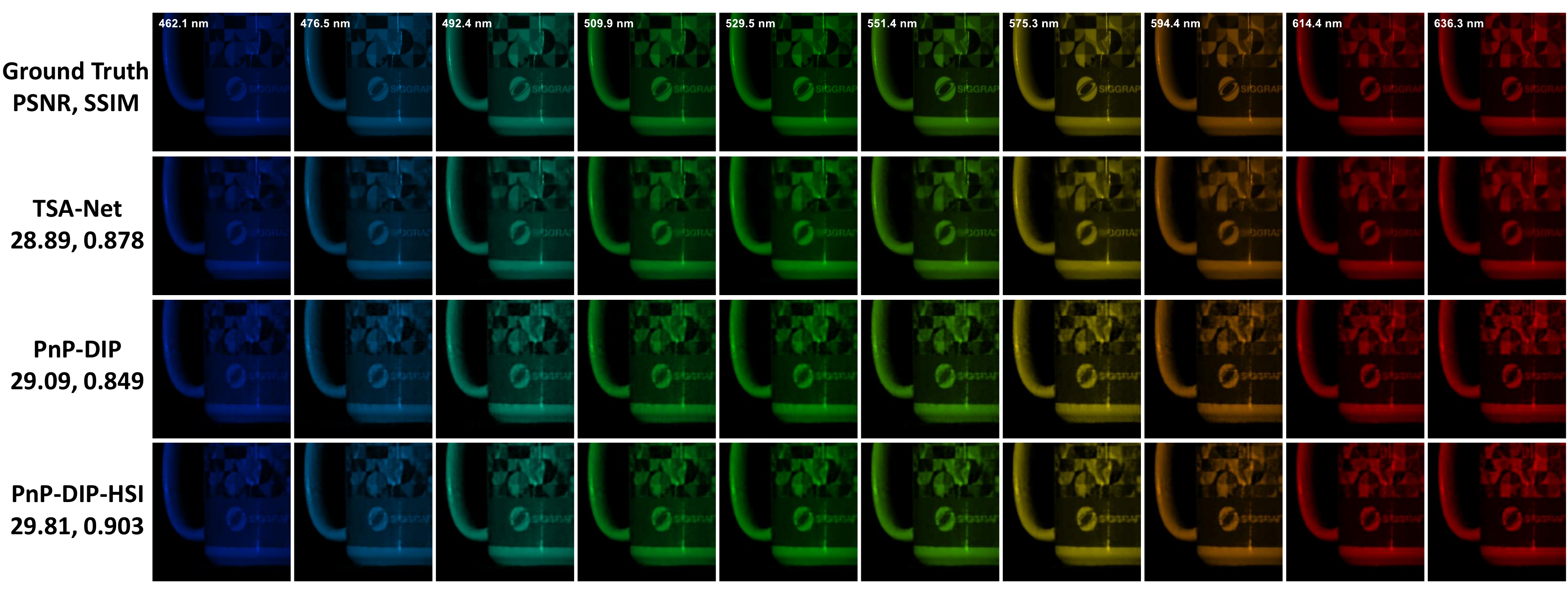}}
\caption{The results of the synthetic data {\em Scene 5} with 10 spectral channels reconstructed by TSA-Net and the proposed PnP-DIP and PnP-DIP-HSI.}
\label{fig:sm7}
\end{figure*}

\begin{figure*}[htbp!]
\centering
\renewcommand\thefigure{M11}
{\includegraphics[width=.98\linewidth]{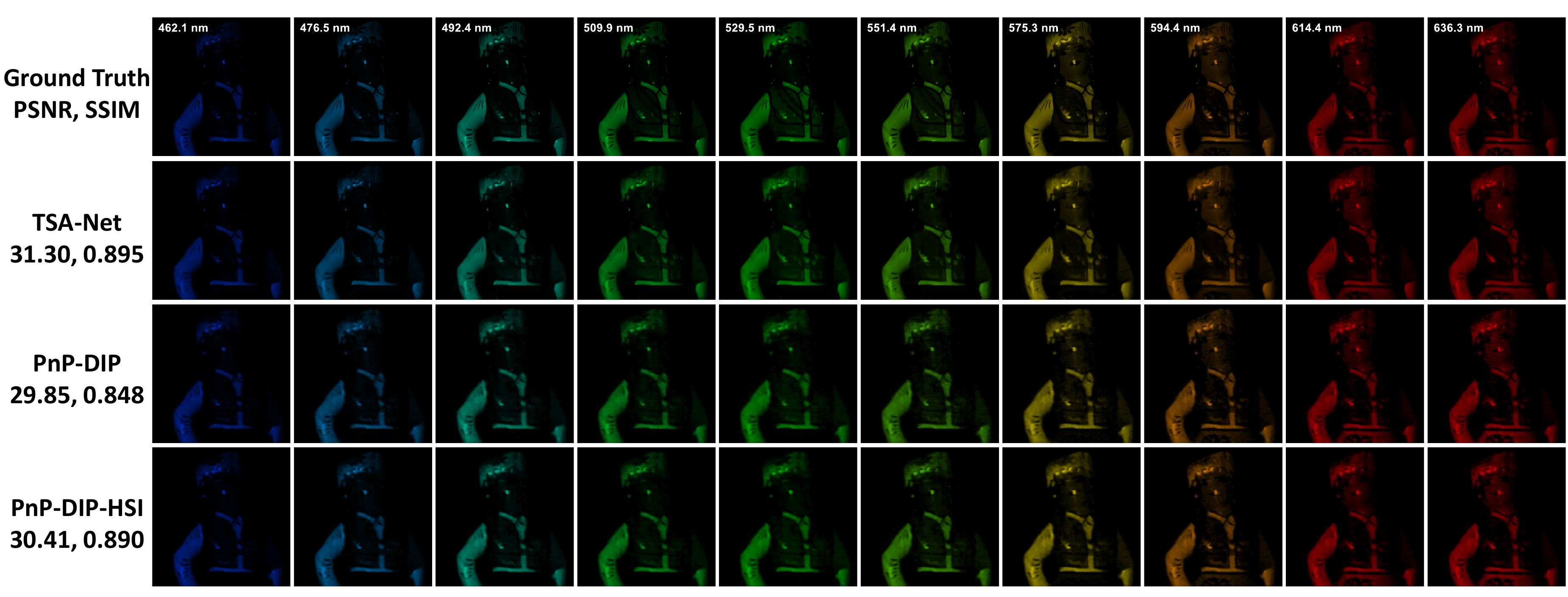}}
\caption{The results of the synthetic data {\em Scene 6} with 10 spectral channels reconstructed by TSA-Net and the proposed PnP-DIP and PnP-DIP-HSI.}
\label{fig:sm8}
\end{figure*}

\begin{figure*}[htbp!]
\centering
\renewcommand\thefigure{M12}
{\includegraphics[width=.98\linewidth]{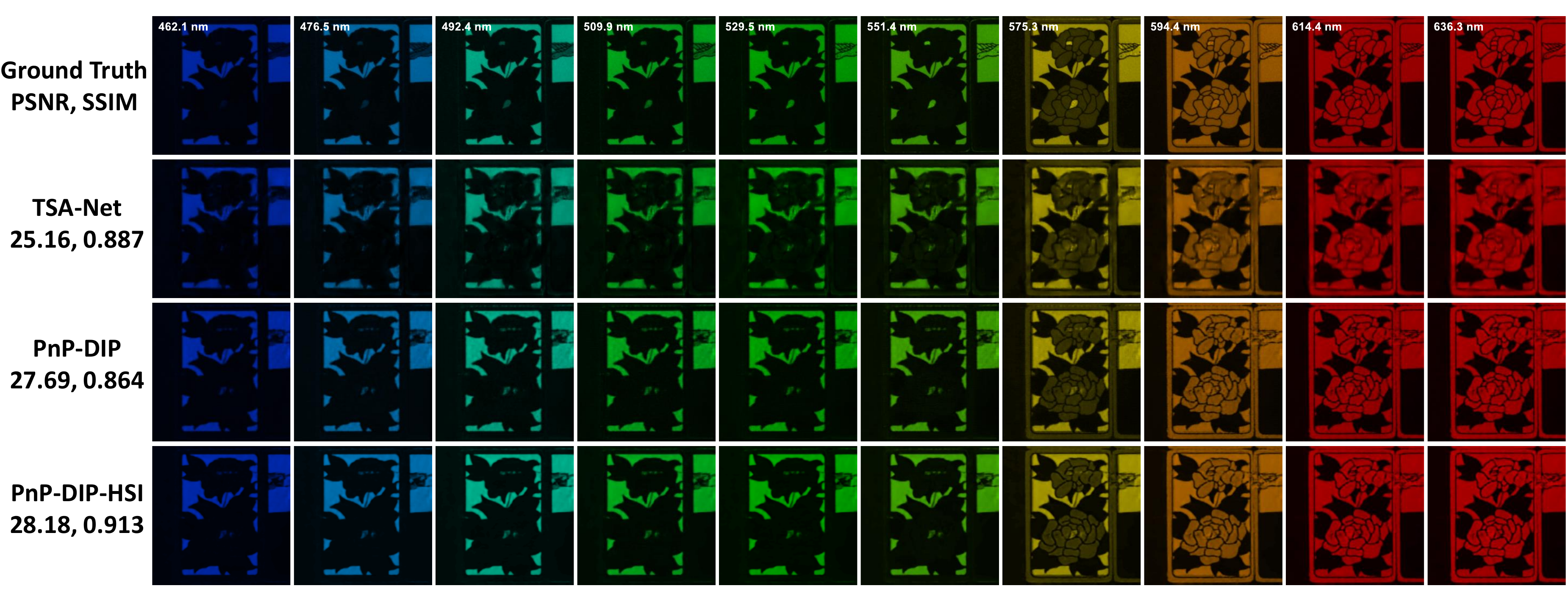}}
\caption{The results of the synthetic data {\em Scene 7} with 10 spectral channels reconstructed by TSA-Net and the proposed PnP-DIP and PnP-DIP-HSI.}
\label{fig:sm9}
\end{figure*}

\begin{figure*}[htbp!]
\centering
\renewcommand\thefigure{M13}
{\includegraphics[width=.98\linewidth]{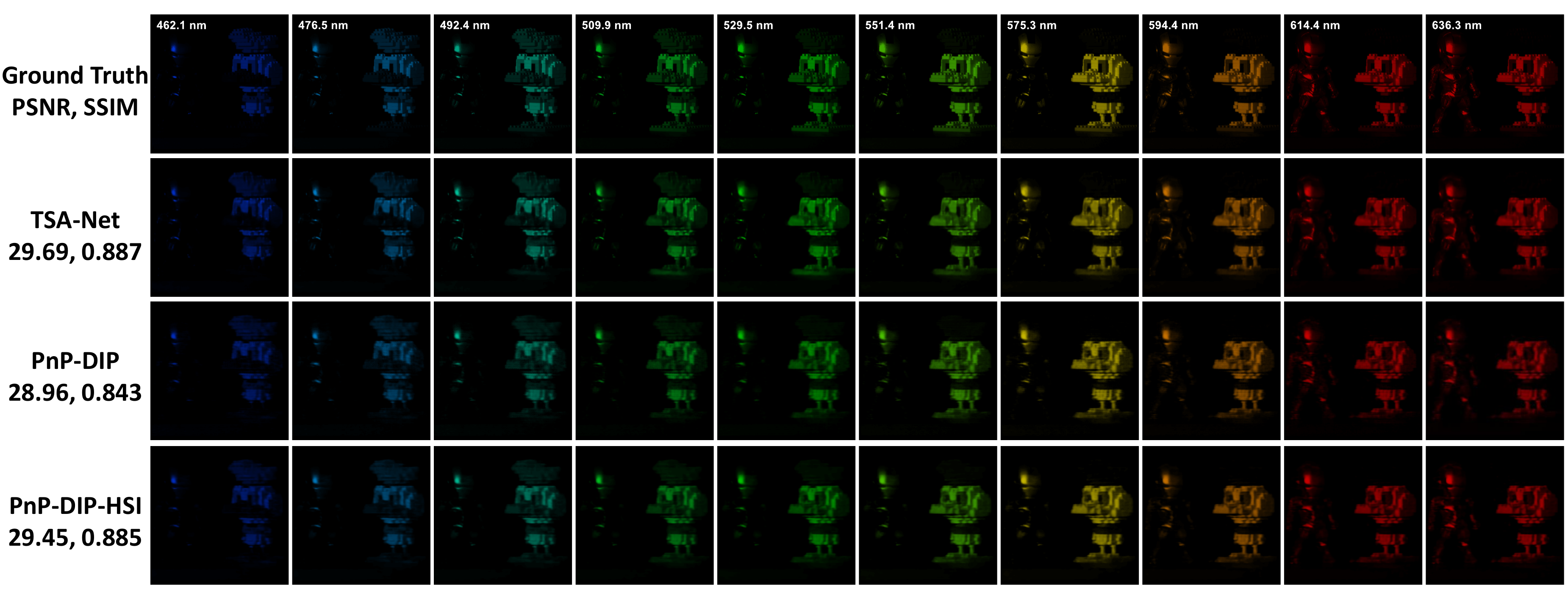}}
\caption{The results of the synthetic data {\em Scene 8} with 10 spectral channels reconstructed by TSA-Net and the proposed PnP-DIP and PnP-DIP-HSI.}
\label{fig:sm10}
\end{figure*}

\begin{figure*}[htbp!]
\centering
\renewcommand\thefigure{M14}
{\includegraphics[width=.98\linewidth]{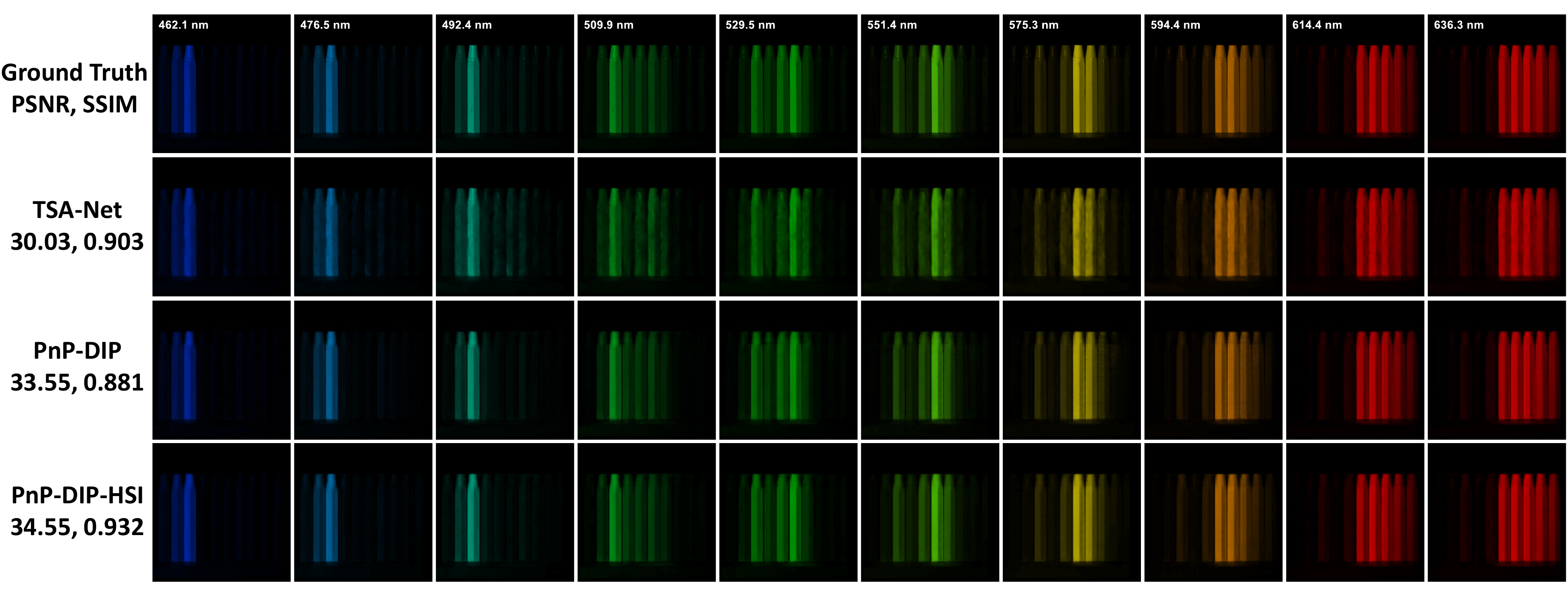}}
\caption{The results of the synthetic data {\em Scene 9} with 10 spectral channels reconstructed by TSA-Net and the proposed PnP-DIP and PnP-DIP-HSI.}
\label{fig:sm11}
\end{figure*}

\begin{figure*}[htbp!]
\centering
\renewcommand\thefigure{M15}
{\includegraphics[width=.98\linewidth]{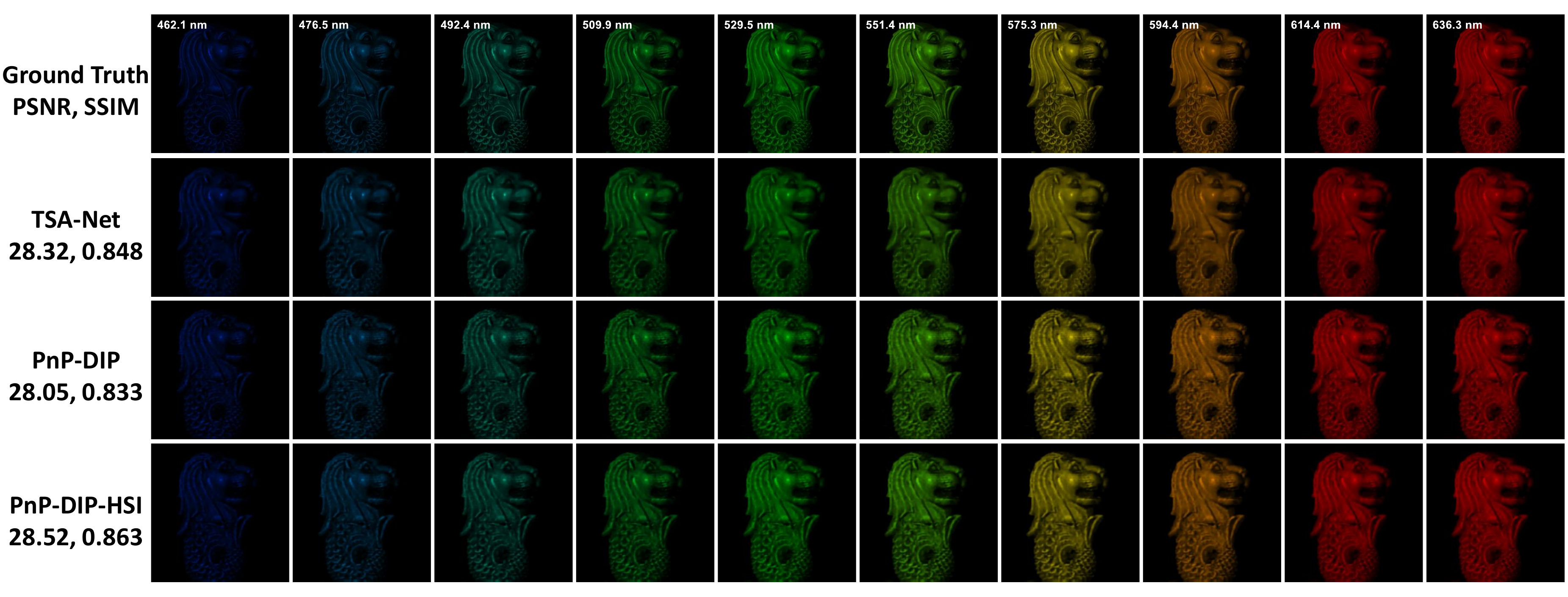}}
\caption{The results of the synthetic data {\em Scene 10} with 10 spectral channels reconstructed by TSA-Net and the proposed PnP-DIP and PnP-DIP-HSI.}
\label{fig:sm12}
\vspace{-1mm}
\end{figure*}

Fig.~\ref{fig:sm3}-\ref{fig:sm12} show the reconstructed results of the synthetic data with 10 out of 28 spectral channels.
We compare the proposed self-supervised method (PnP-DIP) and the method using HSI prior (PnP-DIP-HSI) with the state of the art supervised algorithm (TSA-Net~\cite{Meng20ECCV_TSAnet}) and list the corresponding PSNR and SSIM.

\subsection{Results on the Real Data}
\begin{figure*}[htbp!]
\centering
\renewcommand\thefigure{M16}
{\includegraphics[width=.98\linewidth]{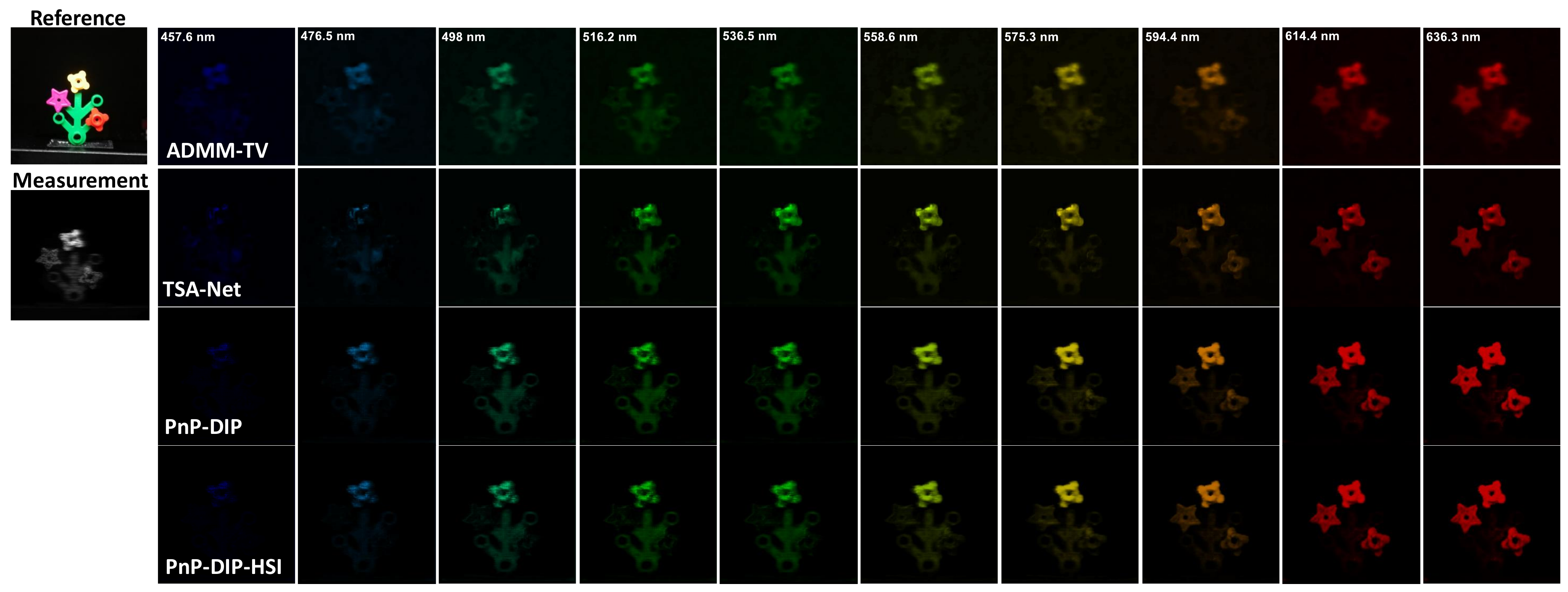}}
\caption{The results of the real data {\em Lego plant} with 10 spectral channels reconstructed by ADMM-TV, TSA-Net and the proposed PnP-DIP and PnP-DIP-HSI.}
\label{fig:sm13}
\vspace{-2mm}
\end{figure*}

\begin{figure*}[htbp!]
\centering
\renewcommand\thefigure{M17}
{\includegraphics[width=.98\linewidth]{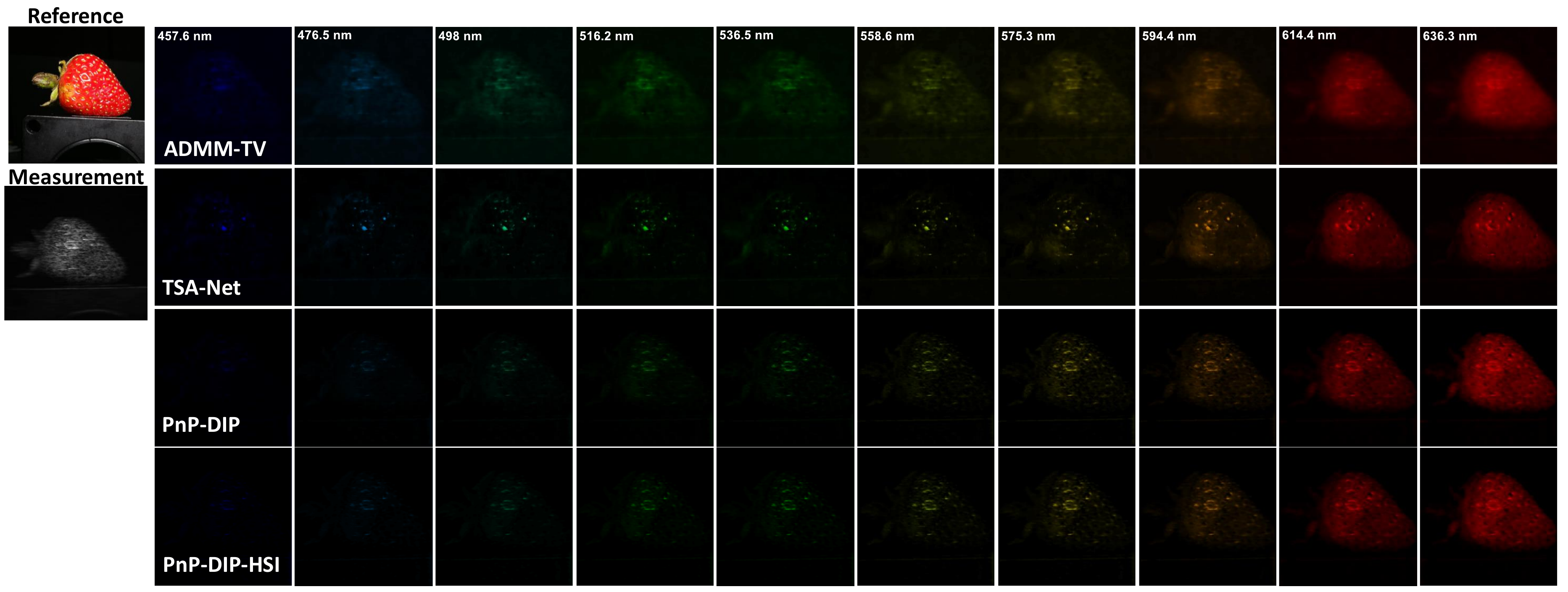}}
\caption{The results of the real data {\em Strawberry} with 10 spectral channels reconstructed by ADMM-TV, TSA-Net and the proposed PnP-DIP and PnP-DIP-HSI.}
\label{fig:sm14}
\vspace{-2mm}
\end{figure*}

\begin{figure*}[htbp!]
\centering
\renewcommand\thefigure{M18}
{\includegraphics[width=.98\linewidth]{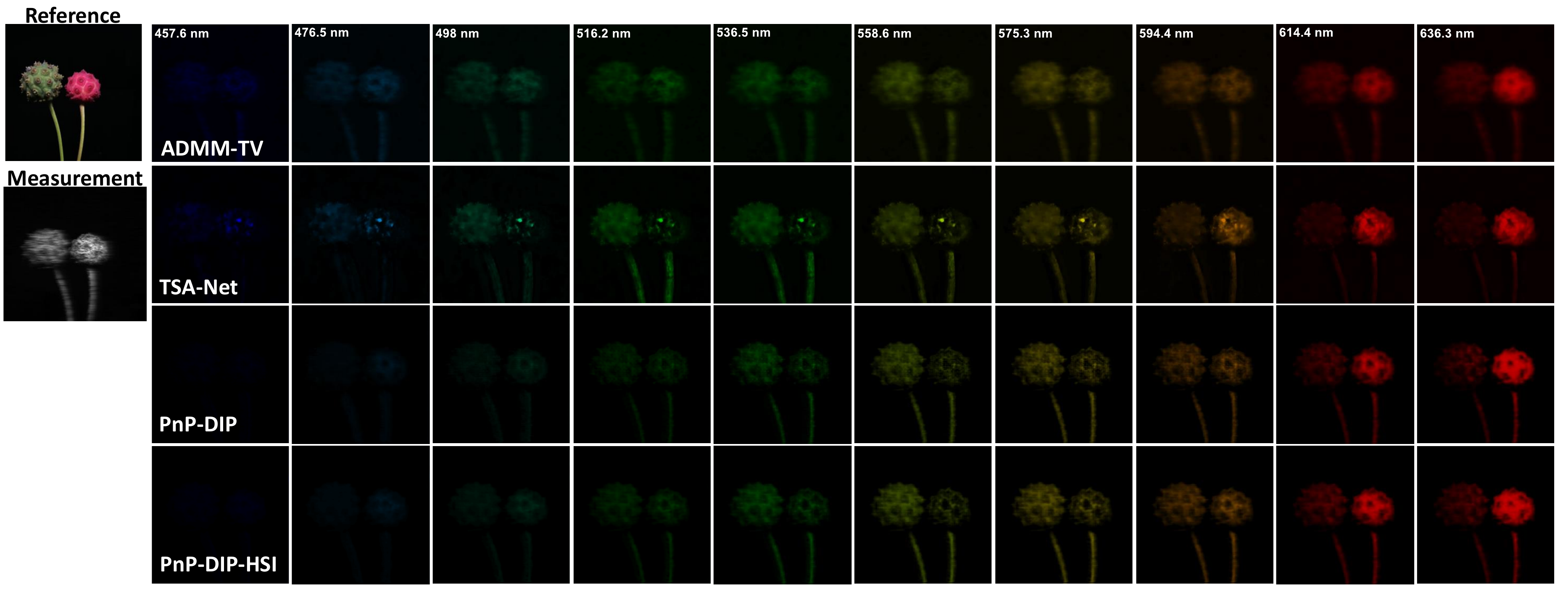}}
\caption{The results of the real data {\em Plant} with 10 spectral channels reconstructed by ADMM-TV, TSA-Net and the proposed PnP-DIP and PnP-DIP-HSI.}
\label{fig:sm15}
\end{figure*}

\begin{figure*}[htbp!]
\centering
\renewcommand\thefigure{M19}
{\includegraphics[width=.98\linewidth]{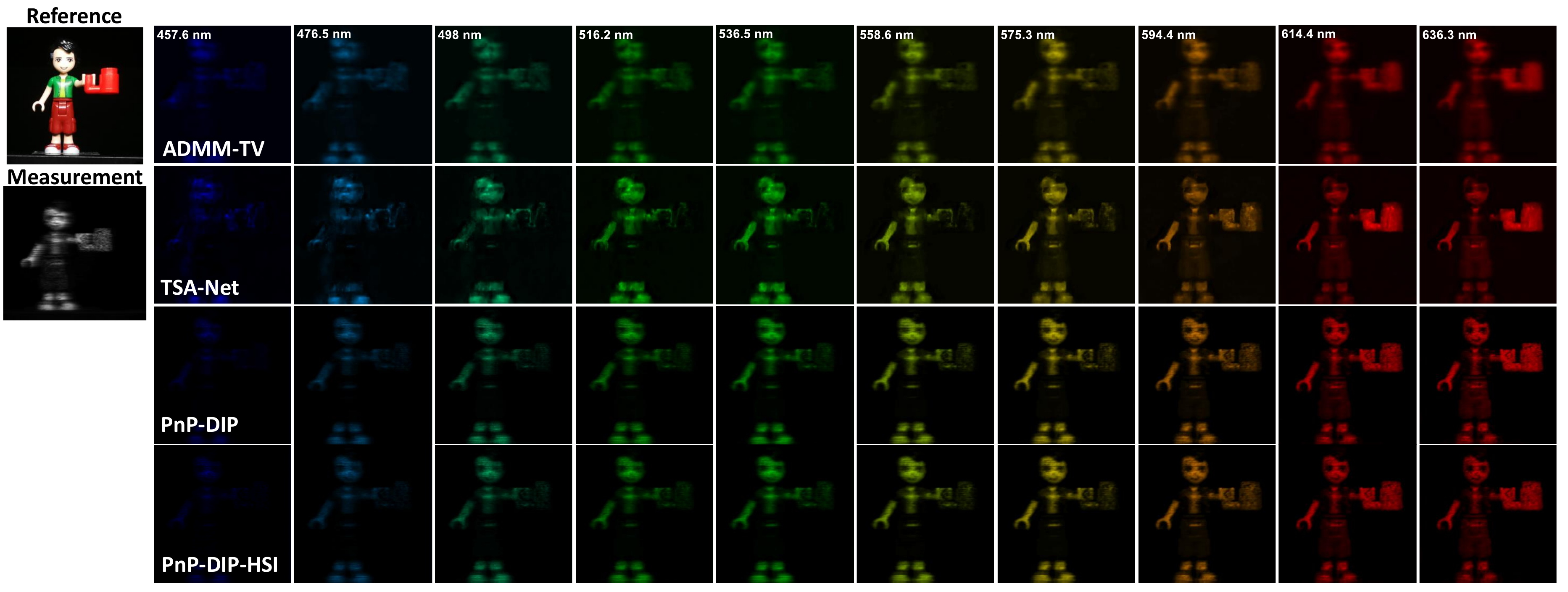}}
\caption{The results of the real data {\em Lego man} with 10 spectral channels reconstructed by ADMM-TV, TSA-Net and the proposed PnP-DIP and PnP-DIP-HSI.}
\label{fig:sm16}
\end{figure*}

\noindent \textbf{CASSI Data Set 1}
We show more results on the datasets captured by the recently bulit CASSI system in~\cite{Meng20ECCV_TSAnet}.
The 2D measurements have a spatial size of $550\times604$, and the recovered spectral cube contains 28 spectral channels with the size of $550\times550$. The specific wavelengths are \{453.3, 457.6, 462.1, 466.8, 471.6, 476.5, 481.6, 486.9, 492.4, 498.0, 503.9, 509.9, 516.2, 522.7, 529.5, 536.5, 543.8, 551.4, 558.6, 567.5, 575.3, 584.3, 594.4, 604.2, 614.4, 625.1, 636.3, 648.1\}nm. 
Fig. ~\ref{fig:sm13}-\ref{fig:sm16} show the reconstructed results of the 4 scenes with 10 out of 28 spectral channels. We compare the proposed self-supervised method (PnP-DIP) and the method using HSI prior (PnP-DIP-HSI) with ADMM-TV and the supervised algorithm TSA-Net~\cite{Meng20ECCV_TSAnet}.

\begin{figure*}[htbp!]
\centering
\renewcommand\thefigure{M20}
{\includegraphics[width=.98\linewidth]{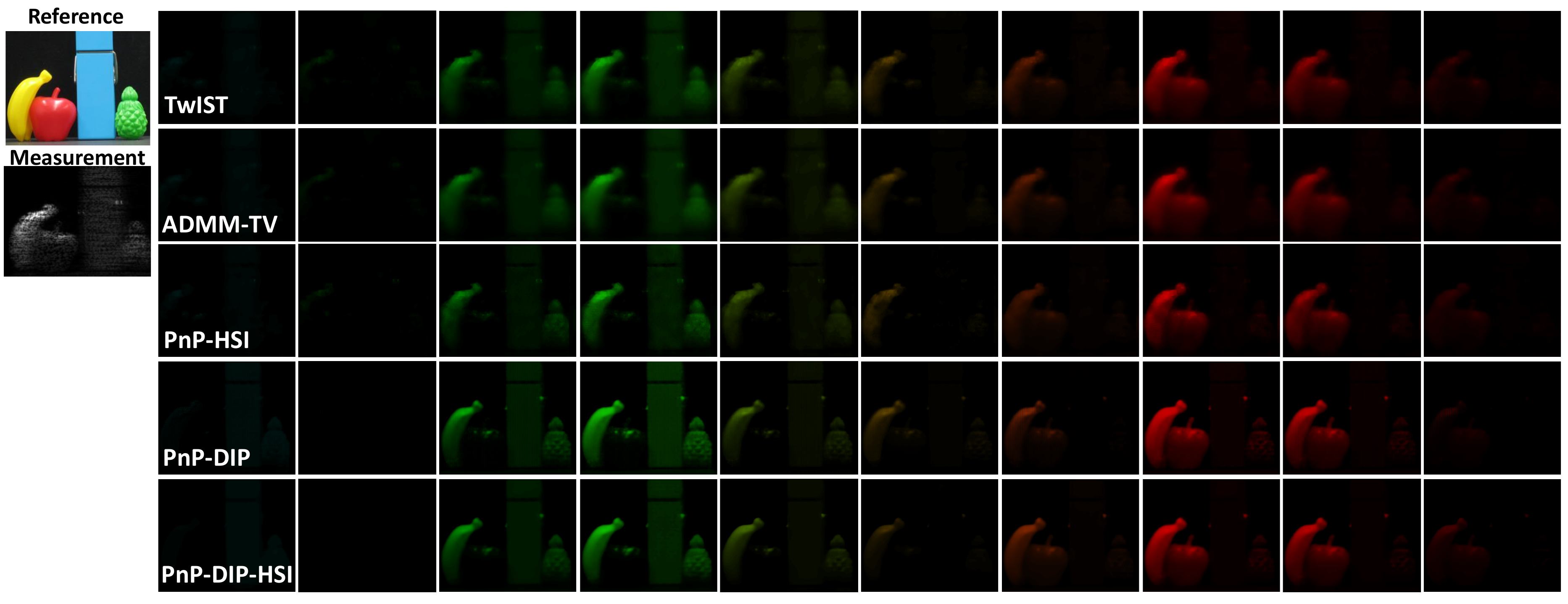}}
\caption{The results of the real data {\em Object} with 10 spectral channels reconstructed by TwIST, ADMM-TV, deep PnP method (PnP-HSI) and the proposed PnP-DIP and PnP-DIP-HSI.}
\label{fig:sm17}
\end{figure*}

\noindent \textbf{CASSI Data Set 2}
We show more results on the datasets captured by the original CASSI system~\cite{kittle2010multiframe}. The reconstructed spectral image contains 33 spectral channels with the size of $210\times256$.
The specific wavelengths are \{454, 458, 462, 465, 468, 472, 475, 479, 483, 487, 491, 496, 500, 505, 509, 514, 520, 525, 531, 537, 543, 549, 556, 564, 571, 579, 587, 596, 605, 615, 626, 637, 650\}nm. 
Fig.~\ref{fig:sm17} show the reconstructed results of the data {\em Object} with 10 out of 33 spectral channels. We compare the proposed self-supervised method (PnP-DIP) and the method using HSI prior (PnP-DIP-HSI) with Twist~\cite{Bioucas-Dias2007TwIST}, ADMM-TV and the deep PnP method (PnP-HSI)~\cite{Zheng20_PRJ_PnP-CASSI}.

\begin{figure*}[htbp!]
\centering
\renewcommand\thefigure{M21}
{\includegraphics[width=.98\linewidth]{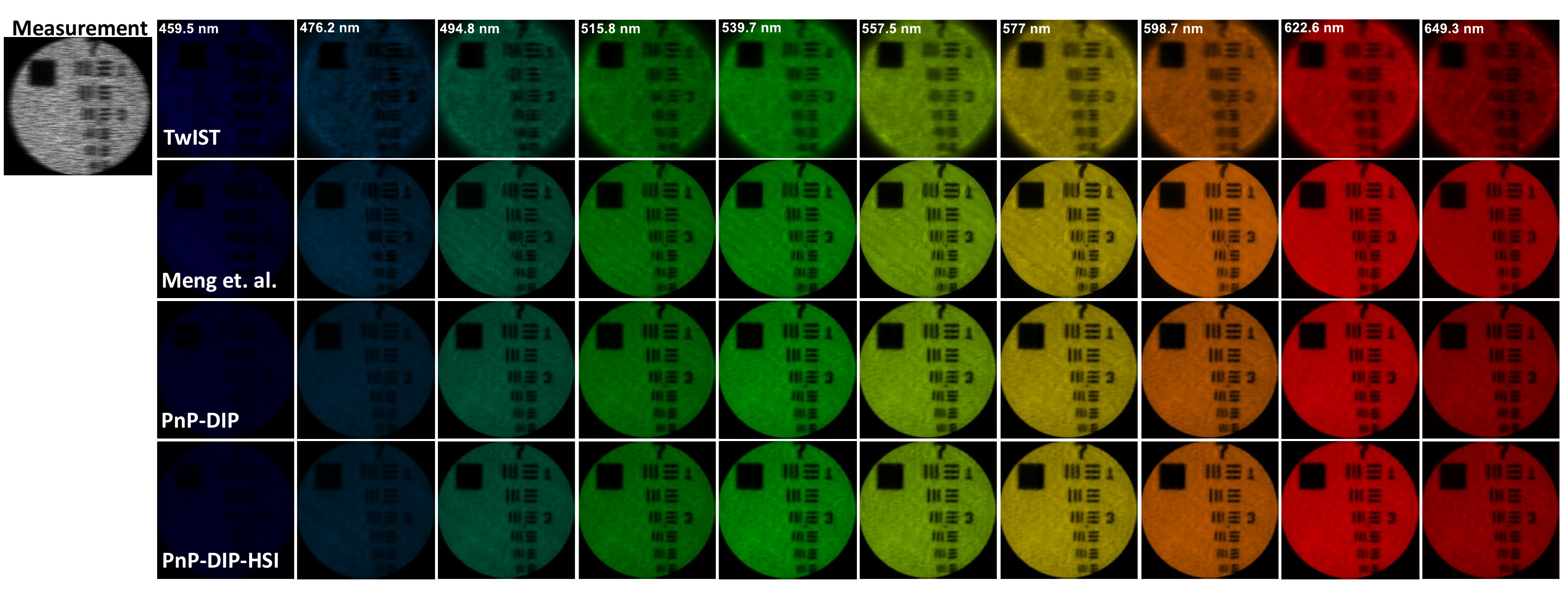}}
\caption{The results of the real data {\em Resolution target} with 10 spectral channels reconstructed by TwIST, a supervised deep neural network and the proposed PnP-DIP and PnP-DIP-HSI.}
\label{fig:sm18}
\end{figure*}

\begin{figure*}[htbp!]
\centering
\renewcommand\thefigure{M22}
{\includegraphics[width=.98\linewidth]{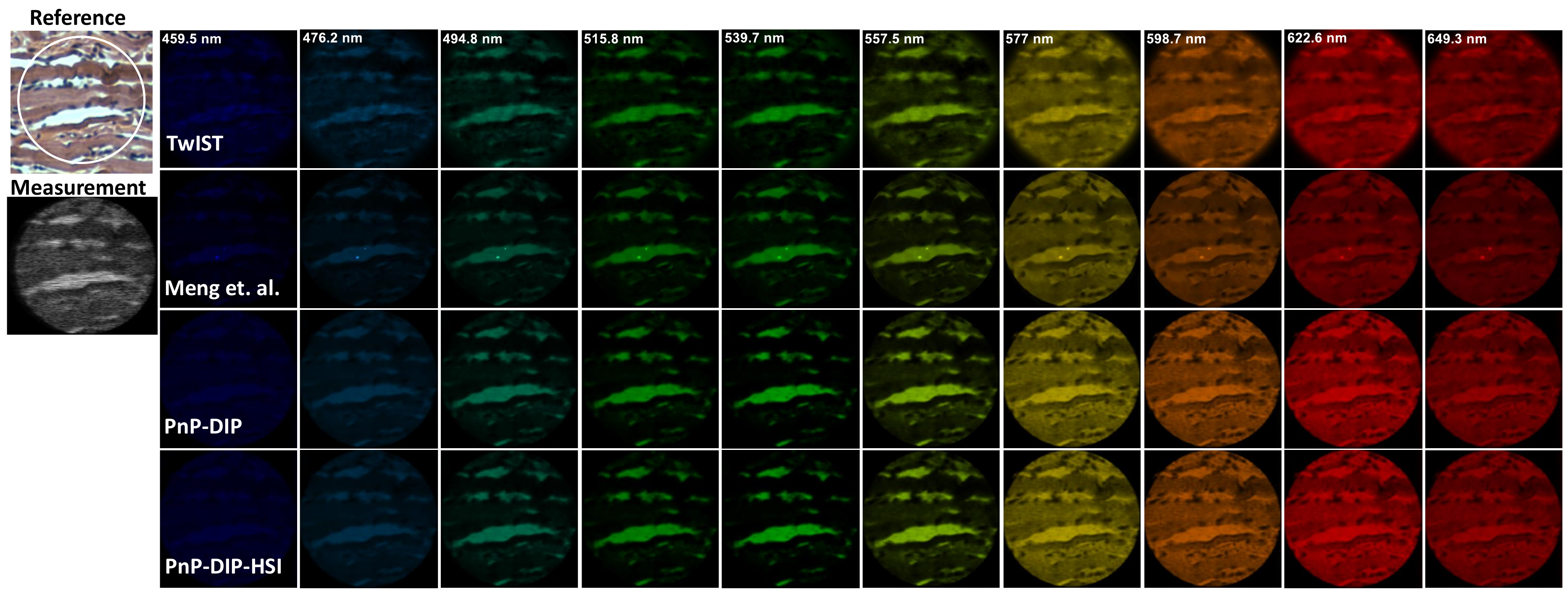}}
\caption{The results of the real data {\em Dog olfactory membrane section 1} with 10 spectral channels reconstructed by TwIST, a supervised deep neural network and the proposed PnP-DIP and PnP-DIP-HSI.}
\label{fig:sm19}
\end{figure*}

\begin{figure*}[htbp!]
\centering
\renewcommand\thefigure{M23}
{\includegraphics[width=.98\linewidth]{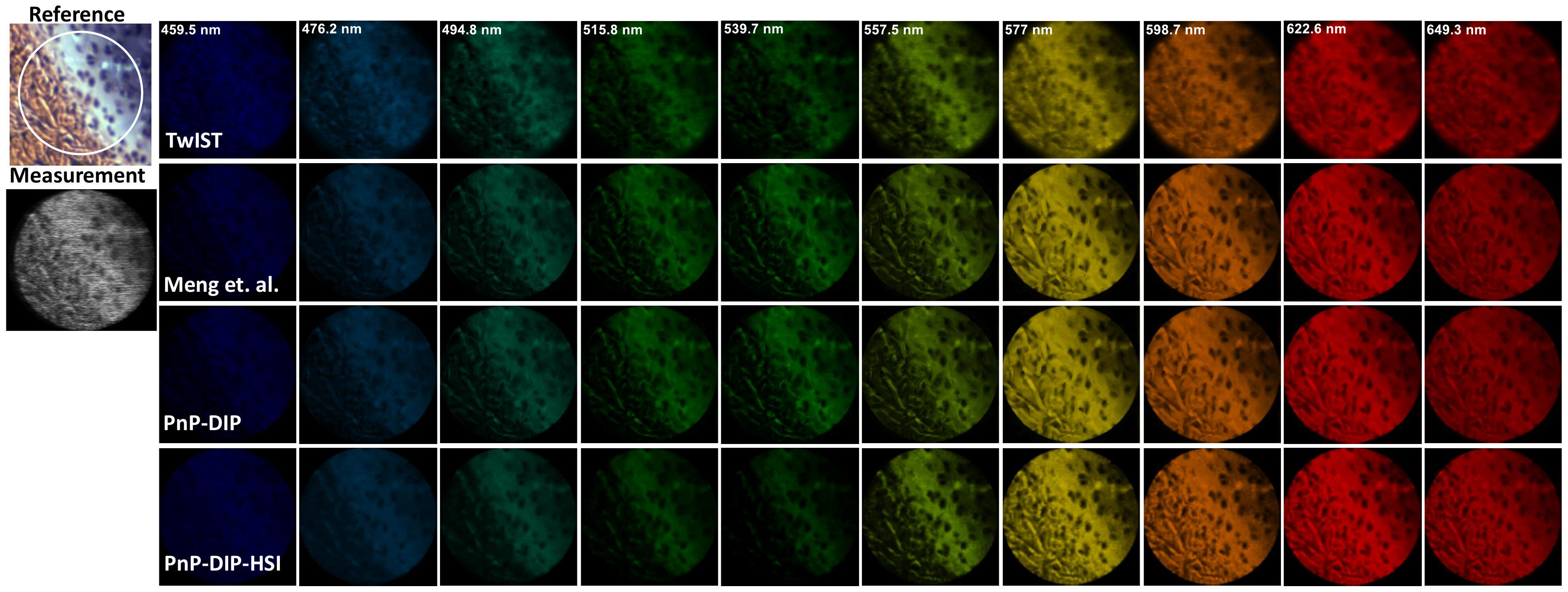}}
\caption{The results of the real data {\em Dog olfactory membrane section 2} with 10 spectral channels reconstructed by TwIST, a supervised deep neural network and the proposed PnP-DIP and PnP-DIP-HSI.}
\label{fig:sm20}
\end{figure*}

\begin{figure*}[htbp!]
\centering
\renewcommand\thefigure{M24}
{\includegraphics[width=.98\linewidth]{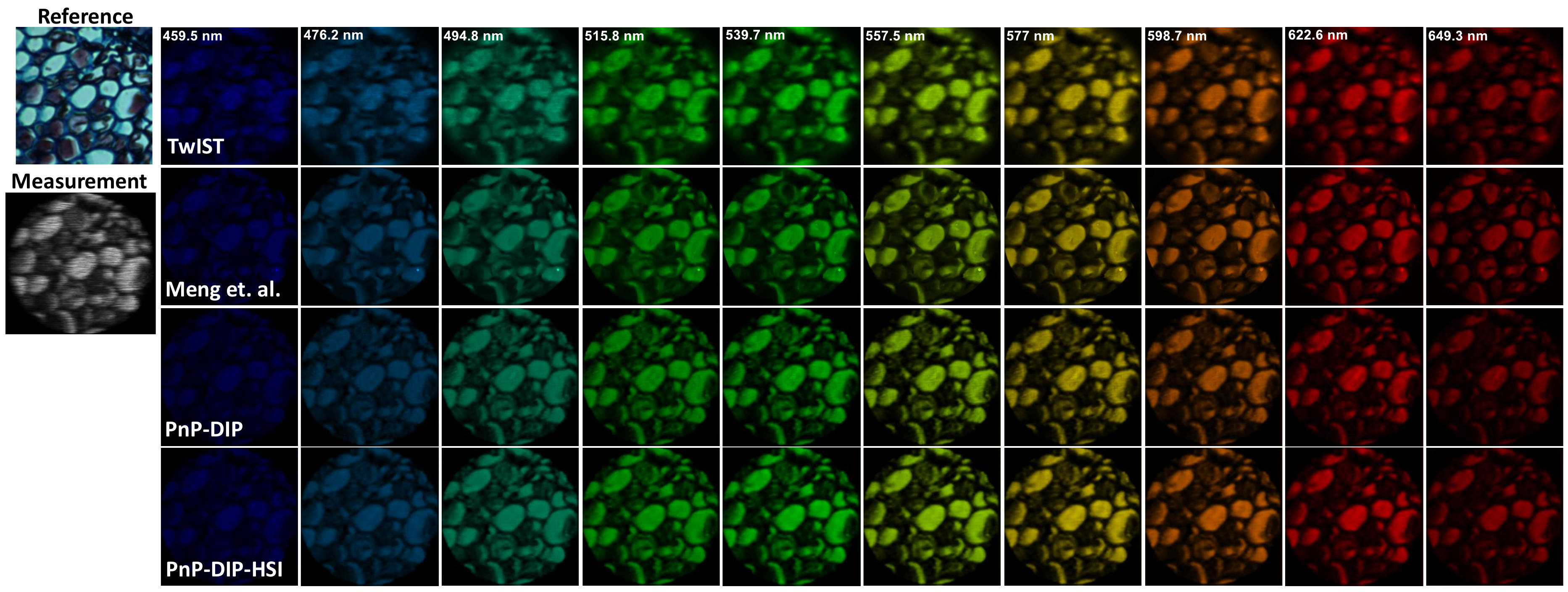}}
\caption{The results of the real data {\em Fern root section} with 10 spectral channels reconstructed by TwIST, a supervised deep neural network and the proposed PnP-DIP and PnP-DIP-HSI.}
\label{fig:sm21}
\end{figure*}

\begin{figure*}[htbp!]
\centering
\renewcommand\thefigure{M25}
{\includegraphics[width=.98\linewidth]{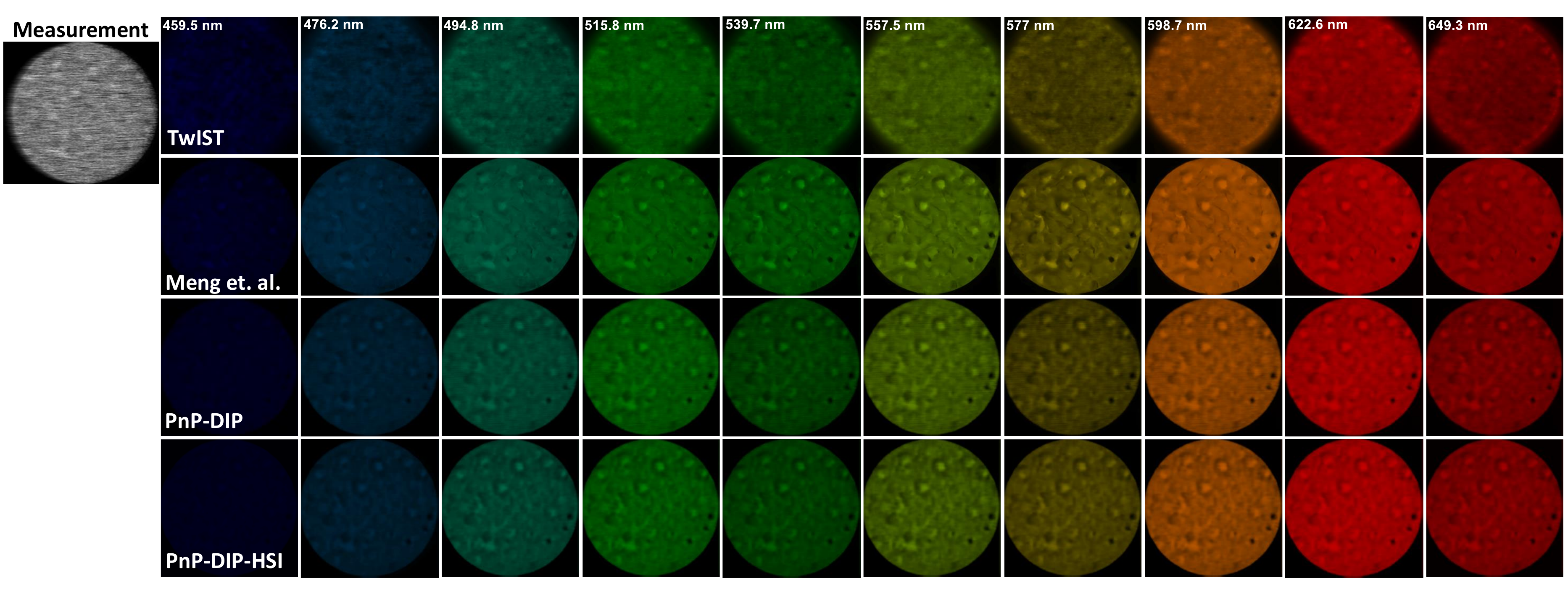}}
\caption{The results of the real data {\em Red blood cell 1} with 10 spectral channels reconstructed by TwIST, a supervised deep neural network and the proposed PnP-DIP and PnP-DIP-HSI.}
\label{fig:sm22}
\end{figure*}

\begin{figure*}[htbp!]
\centering
\renewcommand\thefigure{M26}
{\includegraphics[width=.98\linewidth]{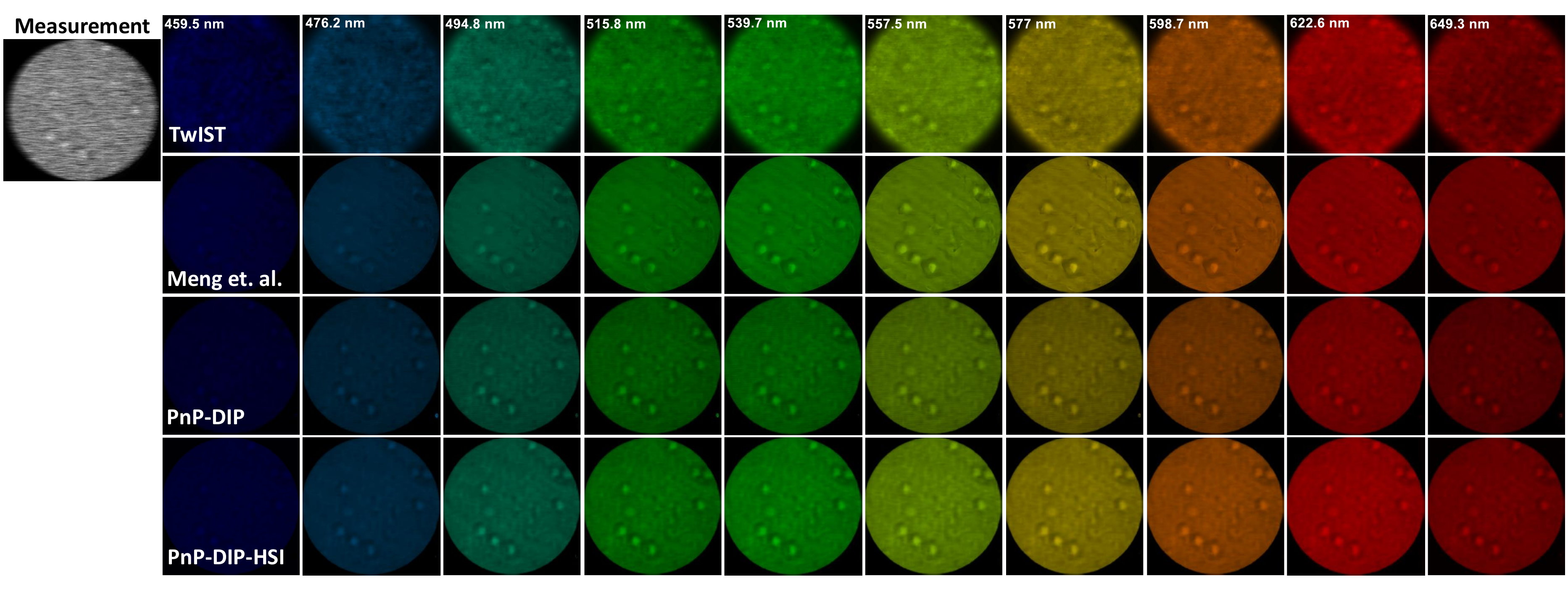}}
\caption{The results of the real data {\em Red blood cell 2} with 10 spectral channels reconstructed by TwIST, a supervised deep neural network and the proposed PnP-DIP and PnP-DIP-HSI.}
\label{fig:sm23}
\end{figure*}

\noindent \textbf{Endomicroscopy Data}
We show more results on the datasets captured by the compressive multispectral endomicroscopy system~\cite{meng2020snapshot}.The captured measurements are of spatial size of $660\times706$, which are used to reconstruct the multispectral endoscopic images with the size of $660\times660\times24$. The specific wavelengths are \{454.4, 459.5, 464.9, 470.5, 476.2, 482.1, 488.4, 494.8, 501.5, 508.5, 515.8, 523.4, 531.4, 539.7, 548.4, 557.5, 567.0, 577.0, 587.6, 598.7, 610.3, 622.6, 635.6, 649.3\}nm. Fig.~\ref{fig:sm18}-\ref{fig:sm23} show the reconstructed results of the 4 scenes with 10 out of 28 spectral channels. We compare the proposed self-supervised method (PnP-DIP) and the method using HSI prior (PnP-DIP-HSI) with TwIST and the supervised deep neural network~\cite{meng2020snapshot}.

\end{document}